\documentclass[twocolumn]{aastex631}
\usepackage{amsmath,amssymb,mhchem}
\usepackage{xspace}
\usepackage{hyperref}
\usepackage{mhchem} 
\usepackage{graphicx}
\usepackage{natbib}

\usepackage{xcolor}
\definecolor{darkgreen}{rgb}{0.0, 0.5, 0.0}

\usepackage{wasysym}

\begin{document}

\title{JWST-TST DREAMS: Secondary Atmosphere Constraints for the \\ Habitable Zone Planet TRAPPIST-1~e}

\shortauthors{Glidden et al.}
\shorttitle{Secondary Atmosphere Constraints for TRAPPIST-1~\lowercase{e}}

\correspondingauthor{Ana Glidden}
\email{aglidden@mit.edu}

\author[0000-0002-5322-2315]{Ana Glidden}
\affiliation{Department of Earth, Atmospheric and Planetary Sciences, Massachusetts Institute of Technology, Cambridge, MA 02139, USA}
\affiliation{Kavli Institute for Astrophysics and Space Research, Massachusetts Institute of Technology, Cambridge, MA 02139, USA}

\author[0000-0002-5147-9053]{Sukrit Ranjan}
\affiliation{Lunar and Planetary Laboratory/Department of Planetary Sciences, University of Arizona, Tucson, AZ 85721, USA}

\author[0000-0002-6892-6948]{Sara Seager}
\affiliation{Department of Earth, Atmospheric and Planetary Sciences, Massachusetts Institute of Technology, Cambridge, MA 02139, USA}
\affiliation{Kavli Institute for Astrophysics and Space Research, Massachusetts Institute of Technology, Cambridge, MA 02139, USA} 
\affiliation{Department of Aeronautics and Astronautics, Massachusetts Institute of Technology, 77 Massachusetts Avenue, Cambridge, MA 02139, USA}

\author[0000-0001-9513-1449]{N\'{e}stor Espinoza}
\affiliation{Space Telescope Science Institute, 3700 San Martin Drive, Baltimore, MD 21218, USA}
\affiliation{William H. Miller III Department of Physics and Astronomy, Johns Hopkins University, Baltimore, MD 21218, USA}

\author[0000-0003-4816-3469]{Ryan J. MacDonald}
\affiliation{School of Physics and Astronomy, University of St Andrews, North Haugh, St Andrews, KY16 9SS, UK}
\affiliation{Department of Astronomy, University of Michigan, 1085 S. University Ave., Ann Arbor, MI 48109, USA}
\affiliation{NHFP Sagan Fellow}

\author[0000-0002-0832-710X]{Natalie H. Allen}
\affiliation{William H. Miller III Department of Physics and Astronomy, Johns Hopkins University, Baltimore, MD 21218, USA}

\author[0000-0003-4835-0619]{Caleb I. Ca\~nas}
\altaffiliation{NASA Postdoctoral Fellow}
\affiliation{NASA Goddard Space Flight Center, Greenbelt, MD 20771, USA}

\author[0000-0001-5878-618X]{David Grant}
\affiliation{University of Bristol, HH Wills Physics Laboratory, Tyndall Avenue, Bristol, UK}

\author[0000-0003-0854-3002]{Am\'{e}lie Gressier}
\affiliation{Space Telescope Science Institute, 3700 San Martin Drive, Baltimore, MD 21218, USA}

\author[0000-0002-7352-7941]{Kevin B. Stevenson}
\affiliation{Johns Hopkins APL, 11100 Johns Hopkins Rd, Laurel, MD 20723, USA}

\author[0000-0003-1240-6844]{Natasha E. Batalha}
\affiliation{NASA Ames Research Center, Moffett Field, CA, 94035, USA}

\author[0000-0002-8507-1304]{Nikole K. Lewis}
\affiliation{Department of Astronomy and Carl Sagan Institute, Cornell University, 122 Sciences Drive, Ithaca, NY 14853, USA}

\author[0000-0002-2508-9211]{Douglas Long}
\affiliation{Space Telescope Science Institute, 3700 San Martin Drive, Baltimore, MD 21218, USA}

\author[0000-0003-4328-3867]{Hannah R. Wakeford}
\affiliation{University of Bristol, HH Wills Physics Laboratory, Tyndall Avenue, Bristol, UK}

\author[0000-0001-8703-7751]{Lili Alderson}
\affiliation{University of Bristol, HH Wills Physics Laboratory, Tyndall Avenue, Bristol, UK}

\author[0000-0002-8211-6538]{Ryan C. Challener}
\affiliation{Department of Astronomy and Carl Sagan Institute, Cornell University, 122 Sciences Drive, Ithaca, NY 14853, USA}

\author[0000-0001-8020-7121]{Knicole Col\'{o}n}
\affiliation{NASA Goddard Space Flight Center, Greenbelt, MD 20771, USA}

\author[0000-0001-5732-8531]{Jingcheng Huang}
\affiliation{Department of Earth, Atmospheric and Planetary Sciences, Massachusetts Institute of Technology, Cambridge, MA 02139, USA}

\author[0000-0003-0525-9647]{Zifan Lin}
\affiliation{Department of Earth, Atmospheric and Planetary Sciences, Massachusetts Institute of Technology, Cambridge, MA 02139, USA}

\author[0000-0002-2457-272X]{Dana R. Louie}
\affiliation{Catholic University of America, Department of Physics, Washington, DC, 20064, USA}
\affiliation{Exoplanets and Stellar Astrophysics Laboratory (Code 667), NASA Goddard Space Flight Center, Greenbelt, MD 20771, USA}
\affiliation{Center for Research and Exploration in Space Science and Technology II, NASA/GSFC, Greenbelt, MD 20771, USA}

\author[0000-0003-0814-7923]{Elijah Mullens}
\affiliation{Department of Astronomy and Carl Sagan Institute, Cornell University, 122 Sciences Drive, Ithaca, NY 14853, USA}

\author[0000-0001-7393-2368]{Kristin S. Sotzen}
\affiliation{JHU Applied Physics Laboratory, 11100 Johns Hopkins Rd, Laurel, MD 20723}

\author[0000-0003-3305-6281]{Jeff A. Valenti}
\affiliation{Space Telescope Science Institute, 3700 San Martin Drive, Baltimore, MD 21218, USA}

\author[0000-0002-2643-6836]{Daniel Valentine}
\affiliation{University of Bristol, HH Wills Physics Laboratory, Tyndall Avenue, Bristol, UK}

\author{Mark Clampin}
\affiliation{NASA Headquarters, 300 E Street SW, Washington, DC 20546, USA}

\author{C. Matt Mountain}
\affiliation{Association of Universities for Research in Astronomy, 1331 Pennsylvania Avenue NW Suite 1475, Washington, DC 20004, USA}

\author[0000-0002-3191-8151]{Marshall Perrin}
\affiliation{Space Telescope Science Institute, 3700 San Martin Drive, Baltimore, MD 21218, USA}

\author[0000-0001-7827-7825]{Roeland P. van der Marel}
\affiliation{Space Telescope Science Institute, 3700 San Martin Drive, Baltimore, MD 21218, USA}
\affiliation{William H. Miller III Department of Physics and Astronomy, Johns Hopkins University, Baltimore, MD 21218, USA}

\begin{abstract}

The TRAPPIST-1 system offers one of the best opportunities to characterize temperate terrestrial planets beyond our own solar system. Within the TRAPPIST-1 system, planet e stands out as highly likely to sustain surface liquid water if it possesses an atmosphere. Recently, we reported the first JWST/NIRSpec PRISM transmission spectra of TRAPPIST-1~e, revealing significant stellar contamination, which varied between the four visits. Here, we assess the range of planetary atmospheres consistent with our transmission spectrum. We explore a wide range of atmospheric scenarios via a hierarchy of forward modeling and retrievals. We do not obtain strong evidence for or against an atmosphere. Our results weakly disfavor CO$_2$-rich atmospheres for pressures corresponding to the surface of Venus and Mars and the cloud tops of Venus at 2$\sigma$. We exclude H$_2$-rich atmospheres containing CO$_2$ and CH$_4$ in agreement with past work, but find that higher mean molecular weight, N$_2$-rich atmospheres with trace CO$_2$ and CH$_4$ are permitted by the data. Both a bare rock and N$_2$-rich atmospheric scenario provide adequate fits to the data, but do not fully explain all features, which may be due to either uncorrected stellar contamination or atmospheric signals. Ongoing JWST observations of TRAPPIST-1~e, exploiting consecutive transits with TRAPPIST-1~b, will offer stronger constraints via a more effective stellar contamination correction. The present work is part of the JWST Telescope Scientist Team (JWST-TST) Guaranteed Time Observations, which is performing a Deep Reconnaissance of Exoplanet Atmospheres through Multi-instrument Spectroscopy (DREAMS).

\end{abstract}


\section{Introduction} \label{intro}

The search for life on exoplanets focuses on atmospheric characterization of temperate rocky worlds, motivated by the potential of such worlds to host global biospheres capable of detectably modifying their atmospheres \citep{Sagan1993, Schwieterman2018}. Ground- and space-based survey telescopes such as TRAPPIST, SPECULOOS, Kepler, and TESS have uncovered thousands of planets, with many orbiting some of our nearest stellar neighbors. However, most of these worlds are unlikely to be conventionally habitable (i.e. hosting surface liquid water). From orbital parameters, we can exclude planets that are either too cool or too hot to allow liquid water on their surface, assuming an Earth-like carbonate-silicate cycle regulating planetary temperature (e.g., \citealt{Kasting1993HabitableStars, Kopparapu2013HabitableEstimates}). In addition, for liquid water to be stable at the surface, the planet must host an atmosphere.  

Of all known exoplanetary systems, TRAPPIST-1 presents the greatest number of potentially habitable rocky worlds with detectable atmospheres \citep{Gillon2017}. The seven rocky planets are all roughly Earth-sized (R$_{p}=0.755-1.129$ R$_{\oplus}$; \citealt{Agol2021}), Earth-mass (M$_{p}=0.326-1.374$ M$_{\oplus}$; \citealt{Agol2021}), and temperate ($T_{eq}=397.6-171.7$ K; \citealt{Ducrot2020}), with TRAPPIST-1~e the most favorably located within the habitable zone. The small size (R$_{\odot}=0.1192\pm0.0013$; \citealt{Agol2021}) and close proximity of the ultracool M dwarf host star renders the TRAPPIST-1 system exceptionally favorable for transmission spectroscopy, which has motivated numerous theoretical and observational atmospheric studies (e.g., \citealt{Morley2017, Fauchez2019, Lustig-Yaeger2019, Mikal-Evans2022, Krissansen-Totton2024ThePlanets}).

The range of possible atmospheric scenarios previously considered for the TRAPPIST-1 planets includes CO$_2$ and N$_2$-dominated atmospheres as seen on Mars, Venus, and Earth; CH$_4$-rich atmospheres, motivated by Titan; and H$_2$-rich atmospheres \citep{deWit2018, Turbet2018}.  Several thermal and nonthermal escape processes are likely at play in their atmospheres (see \citealt{Gronoff2020, Turbet2020} for a review). The intense UV and X-ray irradiation the TRAPPIST-1 planets have received from their host star---in particular during its most active, early stage---may have been enough to strip off their atmospheres entirely \citep[e.g.,][]{VanLooveren2024AiryPlanets, VanLooveren2025}. However, other processes, such as line cooling via CO$_{2}$, could have staved off the loss of a secondary atmosphere \citep{Tian2009, Nakayama2022, Kawamura2024}. The presence of a planetary magnetic field would not necessarily help with atmospheric retention and could instead lead to additional escape \citep{Gunell2018WhyEscape}. Additionally, a secondary atmosphere could have been replenished through processes like volcanism \citep{Dorn2018OutgassingSuper-Earths} and cometary delivery of volatiles \citep{DeNiem2012AtmosphericBombardment}. Some planet formation models also predict that the TRAPPIST-1 planets may have initially formed with a large enough volatile reservoir that they could have survived atmospheric escape \citep[e.g.,][]{Coleman2019, Schoonenberg2019}. There has yet to be a definitive consensus on whether or not the TRAPPIST-1 planets have held onto their secondary atmospheres, though detailed modeling suggests it is possible for TRAPPIST-1~e \citep{Krissansen-Totton2024ThePlanets}. Observations may provide a definitive answer.

Early observations of the TRAPPIST-1 planets with Spitzer and the Hubble Space Telescope (HST) enabled the first constraints on their potential atmospheres \citep[e.g.,][]{deWit2016, deWit2018, Delrez2018}. Using HST Wide Field Camera 3 (WFC3), \citet{deWit2016} observed a combined transit of b and e, finding no spectral features and ruling out a H$_{2}$-dominated atmosphere at 10$\sigma$. \citet{deWit2018} further detected no statistically significant spectral features from 1.1--1.7\,$\micron$ for TRAPPIST-1~d, e, f, and g after observing two transits of each planet with WFC3, including an overlapping transit of g and e. With their data, they were able to rule out H$_{2}$-dominated, cloud-free atmospheres, but the observations were not sensitive to any potential heavier secondary atmospheres. H$_2$-rich envelopes (both cloudy and clear) are also disfavored as the TRAPPIST-1 planets adhere to a single mass-radius relationship \citep{Turbet2020MR, Agol2021}. \citet{Moran2018} extended the analysis of \citet{deWit2018}, arguing that the cloud-top pressures required to reconcile H$_2$-dominated atmospheres with the measured constraints were too high to be physically plausible, thereby favoring nonexistent or higher mean molecular weight ($\mu$) atmospheres. Earlier limits on $\mu$ and clouds are likely overly optimistic as they do not fully consider the impact of stellar contamination, which has proven to be a significant bottleneck to reaching the precision necessary to be sensitive to high-$\mu$, secondary atmospheres \citep{Rackham2018, RackhamEspinoza2023, Lim2023}. 

Observations with JWST offer the promise of detecting secondary atmospheres on the TRAPPIST-1 planets due to JWST's greater sensitivity, spectral resolution, and significantly wider wavelength coverage, which includes prominent features from key molecules such as H$_2$O, CH$_{4}$, and CO$_{2}$ \citep[e.g., see][]{Morley2017, Lustig-Yaeger2019, Lin2021}. However, care is needed to correctly separate the stellar and planetary spectral contributions of any observation. To date, JWST has observed a handful of rocky worlds, including planets b and c in the TRAPPIST-1 system \citep[e.g.,][]{Greene2023, Ih2023, Lim2023, Lincowski2023Potential15m, Zieba2023, Radica2025}. However, no terrestrial atmospheres have been detected. TRAPPIST-1~b was observed for two visits with JWST/NIRISS SOSS, finding significant stellar contamination due to unocculted active regions and no evidence for a secondary atmosphere \citep{Lim2023}. The lack of accurate models to explain the time-varying stellar surface inhomogeneities for an ultracool M8V star caused unreconcilable uncertainties an order of magnitude larger than observation precision alone. Consistent with prior observations with HST \citep{deWit2018}, only a cloud-free, H$_2$-dominated atmosphere could be ruled out with the JWST observations. Additionally, secondary eclipse observations of TRAPPIST-1~b at 15\,$\micron$ with MIRI did not find a thick atmosphere nor evidence of the strong CO$_{2}$ feature at that wavelength \citep{Greene2023}. Further modeling work showed that the emission observation still allows for some atmospheric scenarios such as thick atmospheres that lack strong absorbers like CO$_{2}$ \citep{Ih2023} or that a thick CO$_{2}$ atmosphere is still possible if there is an atmospheric temperature inversion such that CO$_{2}$ becomes an emission feature and indistinguishable from a bare-rock scenario \citep{Ducrot2025}. TRAPPIST-1~c has also been observed in transmission for two transits with JWST NIRISS/SOSS, again finding significant stellar contamination, including both unocculted spots and faculae and no atmospheric features \citep{Radica2025}. However, no atmospheric scenarios could be ruled out for TRAPPIST-1~c. Additionally, MIRI eclipse observations ruled out a thick CO$_{2}$ atmosphere, but thin atmospheres remain possible \citep{Lincowski2023Potential15m, Zieba2023}. 

TRAPPIST-1~e is more likely to have retained an atmosphere than the inner planets in its system as it receives less X-ray and ultraviolet emission from its host star than the innermost planets \citep{Krissansen-Totton2023}. Recently, our companion Letter \citep{Espinoza2025} showed that the first JWST observations of TRAPPIST-1~e using JWST/NIRSpec PRISM have significant contamination due to time-varying stellar surface inhomogeneities, consistent with earlier transmission observations of other planets in the system. \cite{Espinoza2025} presented multiple data reductions, showcasing the moderate gains in precision but significant expansion in wavelength space achieved with JWST compared with earlier HST/WFC3 observations. In particular, NIRSpec/PRISM covers wavelength ranges relevant for H$_2$O, CH$_4$, and CO$_2$ detection. Using a Gaussian process (GP), \citet{Espinoza2025} removed some of the time-varying stellar contributions to the observed spectra. Here, we present a detailed analysis of the atmospheric inferences we can make from both the uncorrected and the stellar-contamination-mitigated spectra. Our analysis aims to reveal the range of planetary atmospheres consistent with our four-transit JWST observations of TRAPPIST-1~e.

This Letter is part of a series by the JWST Telescope Scientist Team (JWST-TST)\footnote{\url{https://www.stsci.edu/\~marel/jwsttelsciteam.html}}, which uses Guaranteed Time Observer (GTO) time awarded by NASA in 2003 (PI: M. Mountain) for studies in three different subject areas: (a) transiting exoplanet spectroscopy (lead: N. Lewis); (b) exoplanet and debris disk coronagraphic imaging (lead: M. Perrin); and (c) Local Group proper motion science (lead: R. van der Marel). A common theme of these investigations is the desire to pursue and demonstrate science for the astronomical community at the limits of what is made possible by the exquisite optics and stability of JWST. The present Letter is part of our work on Transiting Exoplanet Spectroscopy, which focuses on detailed exploration of three transiting exoplanets representative of key exoplanet classes: Hot Jupiters (WASP-17b, GTO~1353), Warm Neptunes (HAT-P-26b, GTO~1312), and Temperate Terrestrials (TRAPPIST-1~e, GTO~1331). 

Our Letter is structured as follows. In Section~\ref{sec:datareduction}, we summarize the results of our companion data Letter, \citet{Espinoza2025}. We then analyze our spectra using a hierarchy of increasingly more sophisticated techniques. In Section~\ref{sec:forwardmodeling}, we first compare our data to a flat line (representative of an airless body or high-altitude cloud deck). We then calculate a model-agnostic lower limit of the mean molecular weight of a putative atmosphere for TRAPPIST-1~e. We further evaluate the plausibility of possible atmospheric scenarios by comparing a grid of forward models to our uncorrected data and GP-corrected data. In Section~\ref{sec:jointgpretrievals}, we use the results of the joint stellar contamination-atmospheric retrievals presented in our companion Letter \citep{Espinoza2025} to evaluate atmospheric constraints. Finally, we discuss the implications of our findings and the path forward in Section~\ref{sec:summary}.


\section{JWST Transmission Spectra of TRAPPIST-1~\MakeLowercase{e}} \label{sec:datareduction} 

\begin{figure*}[!ht]
    \centering
    \includegraphics[width=\linewidth]{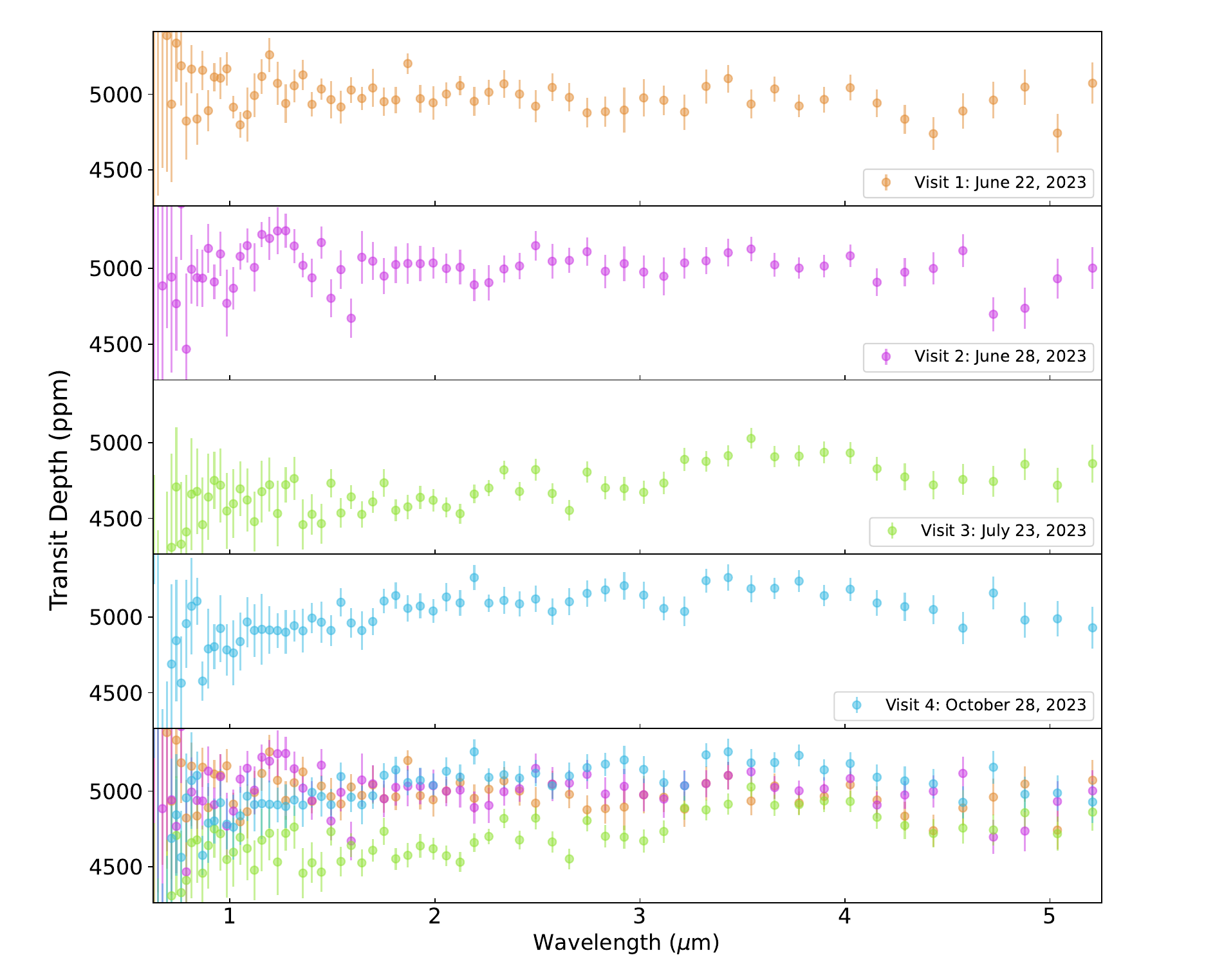}
    \caption{\textbf{JWST NIRSpec PRISM transmission spectra from each of the four visits (uncorrected).} The binned data (R=30) from \cite{Espinoza2025} are shown for each transit separately in the upper four panels and overlaid in the final panel. The x-axis shows the wavelength coverage in microns, while the y-axis shows the transit depth in parts per million (ppm). Each visit exhibits significant transit-to-transit differences in the spectral morphology, indicative of strong contamination from unocculted stellar active regions. The first two visits are the most similar and the least impacted by stellar contamination.}
    \label{fig:transits}
\end{figure*}

In our companion Letter, \cite{Espinoza2025}, we show the first observations of TRAPPIST-1~e with JWST/NIRSpec PRISM. All four of our transits (June 22, June 28, July 23, and October 28, 2023) show evidence for stellar contamination at both the white light curve and spectral level. Visits 1 and 2 were quieter and more mutually similar than Visits 3 and 4. Visit 3 was heavily impacted by stellar contamination, including a large, mid-transit flare. Five reduction methods, including three different data reduction pipelines, were employed. In this work, we make use of the reduction performed by Espinoza using the \texttt{transitspectroscopy} pipeline version 0.4.1 \citep{transitspectroscopy} as it has received additional treatment to mitigate the effects of stellar contamination \citep[][Appendix A.1]{Espinoza2025}. The uncorrected spectra are shown in Figure \ref{fig:transits}.

Making use of traditional stellar models proved unsuccessful at mitigating the impact of stellar contamination \citep[][ Appendix B]{Espinoza2025}. The observed surface inhomogeneities were poorly modeled by combining ``hot" and ``cool" stellar models. Instead, \citet{Espinoza2025} used a GP to model our ignorance in the stellar contribution to the spectra that caused visit-to-visit variations and performed joint stellar and atmospheric retrievals. The GP is only able to model out broad, time-varying trends and does not take into account physical properties of spots, faculae, granulation, etc. Despite these constraints, it provides our best approach at mitigating the impact of stellar activity given the limitations of current stellar models. 

\begin{figure*}[!ht]
    \centering
    \includegraphics[width=\linewidth]{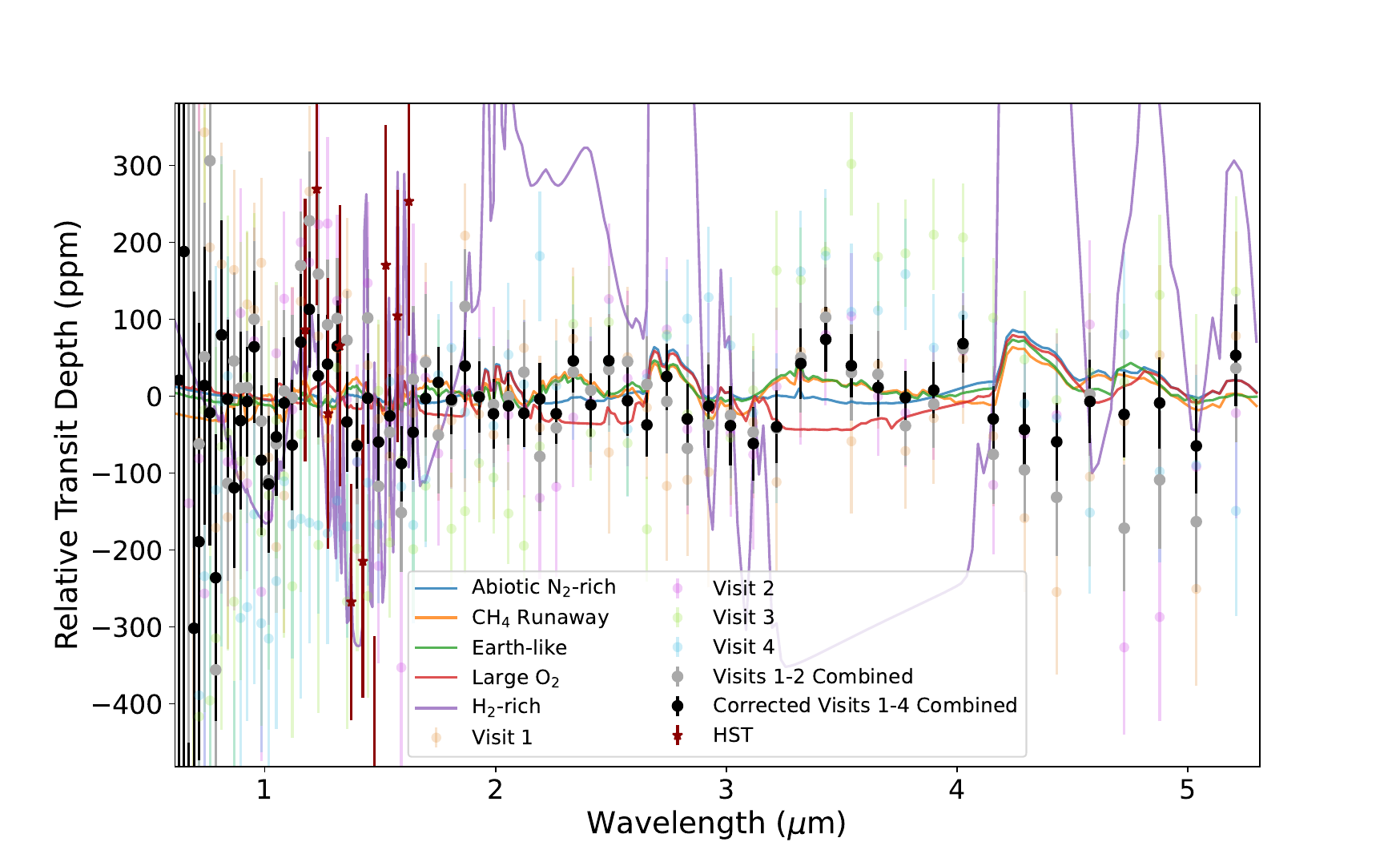}
    \caption{\textbf{JWST NIRSpec PRISM transmission spectra compared with modeled atmospheric scenarios.} The x-axis shows wavelengths in microns, while the y-axis shows the transit depth in parts per million. Each of the individual visits (translucent colored circles with error bars), the combination of Visits 1 and 2 (gray circles with error bars) and the stellar-contamination-corrected four-visit spectrum from \cite{Espinoza2025} (black circles with error bars) are compared to the HST data from \citet{deWit2018} (dark red stars with error bars). An array of forward model spectra for different atmospheric scenarios is overlaid (colored lines). Atmospheric models rich in H$_2$ can be easily ruled out, while many high mean molecular weight scenarios are consistent with the JWST data.}
    \label{fig:sample}
\end{figure*}

We compare both our raw and corrected data in Figure \ref{fig:sample}. Each visit is shown using colored markers, while the combined Visits 1+2 are shown in gray and the GP-corrected Visits 1--4 are shown in black. Several representative forward models are also shown for visual comparison, including a N$_2$-rich, abiotic case; methane runaway; an Earth-like case; a O$_2$-rich case; and a H$_2$-rich case. The O$_2$ and H$_2$-rich cases do not use photochemistry, while the CH$_4$ runaway, N$_2$-rich abiotic, and Earth-like cases do. We assume a dry adiabatic pressure-temperature profile evolving into an isothermal stratosphere \citep{Pierrehumbert2010}. For the cases with photochemistry, we calculate the atmospheric profiles using using {\tt MEAC} \citep{Hu2012, Hu2013, Ranjan2022}. We apply top- and bottom-of-atmosphere chemical boundary conditions (outgassing, surface deposition, escape) using either modern Earth \citep{Hu2012}, abiotic Earth \citep{Ranjan2020}, or photochemical runaway \citep{Ranjan2022} boundary conditions for the Earth-like, N$_2$-rich abiotic, and CH$_4$ runaway cases, respectively. The forward model spectra are generated with {\tt petitRADTRANS} \citep[{\tt pRT};][]{Molliere2019prt}. In Section \ref{sec:forward_model_grid}, we systematically and quantitatively compare our data with a grid of forward models. Additional details about our forward modeling framework are included in Appendix \ref{sec:appendixD}.

In our hierarchy of analysis, we first choose to use the uncorrected spectra, but only the quieter Visits 1 and 2. In Appendix \ref{sec:appendixA}, we detail how we have combined the spectral data from the first two visits. At each stage in our analysis, we then also use the corrected results from the joint stellar and atmospheric retrieval to make further inferences.


\section{Comparisons with Forward Models} \label{sec:forwardmodeling} 

We evaluate how our JWST data can provide additional constraints on the presence and composition of an atmosphere on TRAPPIST-1~e through comparisons with forward models. Our four-transit dataset serves as preliminary reconnaissance of the potential atmosphere of TRAPPIST-1~e, but this exploration is intrinsically limited by the small number of transits, the active nature of the host star, and the lack of satisfactory stellar models. 

As shown in \cite{Espinoza2025}, all of our transits were impacted by stellar activity with the heaviest contamination occurring in our third and fourth observations. Here, we apply a hierarchy of atmospheric model complexity given the wide array of factors (both astrophysical and instrumental) contributing to our observed spectra of TRAPPIST-1~e. This hierarchical approach ensures that we do not over-interpret our data. Despite these limitations, we can place stronger constraints on the potential atmosphere of TRAPPIST-1~e than previously possible from HST.

\subsection{Comparison of Observed Spectra to a Flat Line} \label{Sec:flat_line}

First, we test if our individual visits and combined-visit data are consistent with a flat-line model (i.e., a bare rock or an extremely cloudy atmosphere). We perform a basic $\chi^2$ test as a first approach, noting that the test does not account for correlations within the data. Our later Bayesian analysis takes these correlations into account. In Table~\ref{tab:chisquaredflatline}, we report standard statistics for a frequentist flat-line rejection test applied to our JWST transmission spectra. Included in the table are the degrees of freedom ($\nu$) associated with the $\chi^2$, the reduced $\chi^2$ ($\chi^2_r=\chi^2/\nu$), and the mean precision of the spectral data points. $\nu$ is equal to the number of data points in the sample minus one. We calculate the mean precision as follows, where $\Delta y_i$ is the precision of each spectral data point and N is the number of points:
\begin{equation}
    \overline{\Delta y}=\frac{1}{N}\sum_{i=1}^{N}\Delta y_i
\end{equation}

From these quantities, we calculate the p-value, which is defined as the probability of obtaining a $\chi^2$ statistic at least as extreme as the one observed, under the assumption that the null hypothesis is true. In our context, the null hypothesis states that the data are consistent with a flat-line model. The smaller the p-value, the stronger the evidence against the flat-line model. A common threshold for rejecting the null hypothesis is a p-value $\leq$ 0.05. We use the equation by \citep{BevingtonRobinson2003}:
\begin{equation}
    P(\chi^2 | \nu, H_0)=\int_{\chi^2}^{\infty} \frac{x^{\nu/2 - 1}e^{-x^2/2}}{2^{\nu/2}\Gamma(\nu/2)} dx^{2}
\end{equation}
where $\Gamma$ is the $\Gamma$ function and $\nu$ are the degrees of freedom of the $\chi^{2}$ distribution. Finally, we quantify the statistical significance of how well a flat line can be ruled out, given the data, by calculating the percentage point function (the inverse of the cumulative distribution function). 

We find Visits 1 and 2 are consistent with a flat line from this simple analysis, while a flat line is moderately rejected at 2.4$\sigma$ for Visit 4 and nominally ruled out at 6$\sigma$ for Visit 3. However, since Visits 3 and 4 are strongly contaminated by stellar activity, they are unreliable as atmosphere proxies without extensive mitigation \citep[see][]{Espinoza2025}. The relatively flat individual visit JWST data are consistent with past observations ruling out a H$_2$-dominated atmosphere on TRAPPIST-1~e \citep{deWit2018, Moran2018}, as well as predictions that $\geq 7$ transits are required to detect atmospheric features of a high mean molecular weight secondary atmosphere \citep{Morley2017, Lustig-Yaeger2019}, even when not considering the impact of clouds and stellar contamination. When clouds and/or hazes are included, it is predicted to take up 15 transits to detect CO$_2$ at 3$\sigma$ \citep{Fauchez2019}.

We further find that our GP-corrected Visits 1--4 combined spectrum is consistent with a flat line. However, there is correlated residual noise that the flat-line model does not account for. Using our hierarchy of modeling, we explore the residual undulations through both forward modeling in Section \ref{sec:forward_model_grid} and through a fully Bayesian analysis in Section \ref{sec:jointgpretrievals}.

While the mean precision of a single JWST visit is consistent with an earlier two-visit HST transmission spectrum \citep{deWit2018}, the larger wavelength coverage of the JWST NIRSpec PRISM (1.1--1.7$\micron$ vs. 0.6--5.2$\micron$) allows us to place tighter constraints on the allowable mean molecular weight. Our JWST data cover wavelengths where strong CH$_4$ ($\sim3.3\micron$) and CO$_2$ ($\sim4.3\micron$) features are expected, allowing us to use non-detections of spectral features to constrain properties of any potential atmosphere. We proceed to quantify such atmospheric constraints.

\subsection{Model-agnostic Limits on the \\ Mean Molecular Weight of TRAPPIST-1~e}  \label{sec:mu_results}

We begin with a simple, model-agnostic approach to constrain $\mu$ consistent with TRAPPIST-1~e's JWST transmission spectrum. If TRAPPIST-1~e had a low mean molecular weight atmosphere (e.g., $\mu = 2.3$\,u for a H$_2$+He-dominated atmosphere, where `u' is the atomic mass unit), we would readily observe signatures of molecules like CO$_2$ expected to be present in rocky planet atmospheres \citep[e.g.,][]{Moran2018,Lustig-Yaeger2019}. However, as we do not detect molecular features, our observations can be used to place a lower limit on $\mu$ for a putative TRAPPIST-1~e atmosphere.

\begin{deluxetable}{lcccccc}
    \renewcommand{\arraystretch}{1.1}
    \tabletypesize{\footnotesize}
    \tablecolumns{6} 
    \tablecaption{\textbf{Flat-line model rejection test for TRAPPIST-1~e}}
    \tablehead{Visit & $\chi^{2}$ & $\nu$ & $\chi_{\nu}^{2}$ & $\overline{\Delta y}$ (ppm) & p-value & $N\sigma$} 
    \startdata
    \hline
    \hline
    \#1: June 22, 2023 & 66 & 66 & 0.99 & 226 & 0.49 & 0.69 \\
    \#2: June 28, 2023 & 64 & 66 & 0.97 & 234 & 0.56 & 0.59 \\
    \#3: July 23, 2023 & 156 & 66 & 2.37 & 221 & 0.0 & 5.9\\
    \#4: October 28, 2023 & 93 & 66 & 1.41 & 244 & 0.02 & 2.4\\
    \#1+2 Combined & 67 & 66 & 1.01 & 164 & 0.45 & 0.75\\
    \#1--4 Combined & 44 & 66 & 0.67 & 118 & 0.98 & 0.02 \\
    (decontaminated) & & & & & & \\
    HST & 18.9 & 9 & 2.1 & 171 & 0.03 & 2.2\\
    \enddata
    \tablecomments{JWST data from \cite{Espinoza2025}, HST data from \citet{deWit2018}. $\nu$ is the number of degrees of freedom, $\chi_{\nu}^{2}$ is the reduced chi-squared, $\overline{\Delta y}$ is the mean data precision, $P(\chi^2, \nu)$ is the p-value, and $N\sigma$ is the statistical significance by which a flat line is rejected by the data.
    }
    \label{tab:chisquaredflatline}
\end{deluxetable}

We estimate the limit on $\mu$ as follows. The amplitude of spectral features can be expressed as \citep{Seager2000, Brown2001, Seager2010} 
\begin{equation} \label{eq:amplitude}
    \Delta \delta=\frac{(R_p+nH)^2}{R_*^2}-\frac{R_p^2}{R_*^2} \approx \frac{2nHR_p}{R_*^2}
\end{equation}
where $R_p$ is the planetary radius, $R_*$ is the stellar radius, and the factor of 2 comes from approximating the atmosphere area as an annulus with width $H$, where $H$ is the scale height of the planet's atmosphere given as
\begin{equation} \label{eq:scale_height}
H=\frac{kT}{\mu g}
\end{equation}
where $k$ is the Boltzmann constant; $T$ is the atmospheric temperature; and $g$ is gravitational acceleration. We consider $n=2-5$ \citep{Seager2010, Stevenson2016, Brande2024CloudsExoplanets}.

If there were spectral features present with amplitude $\geq5\times$ the uncertainty on the data, we would detect an atmosphere at 5$\sigma$. Since no such features are present, we can write
\begin{equation}
5 \overline{\Delta y}~>~\Delta \delta \label{eq:ineq}
\end{equation}
Using Equations~\ref{eq:amplitude} and \ref{eq:ineq}, we solve for a lower limit on $\mu$:
\begin{equation}
    \mu>\frac{2 n R_p k T}{5 R_*^2 g \overline{\Delta y}}=\frac{2 n R_p^3 k T}{5 G M_p R_{*}^2\overline{\Delta y}}
\end{equation}
where we have used $g = G M_p / R_p^2$. We can thus use our featureless spectrum to place a limit on $\mu$, assuming a clear-sky atmosphere with atmospheric absorbers present \citep{Moran2018}.

To estimate the uncertainty on $\mu$, $\sigma_\mu$, we propagate the uncertainty on $\overline{\Delta y}$, $\sigma_{\overline{\Delta y}}$, assuming normal errors. We estimate $\sigma_{\overline{\Delta y}}$ via the formula for the standard error of the mean:
\begin{equation}
    \sigma_{\overline{\Delta y}} = \frac{sd(\Delta y)}{\sqrt{N}}
\end{equation}
where $sd(\Delta y)$ is the standard deviation of the uncertainties and $N$ is the number of data points. For context, giant planet like Jupiter, Saturn, and Uranus are typically have $\mu\sim2$u (H$_2$-dominated), while terrestrial solar system planets have $\mu\sim28-44$u (N$_2$ and CO$_2$-dominated, respectively).

We first apply this order-of-magnitude estimate to the data from Visits 1 and 2 as they are less affected by stellar contamination. We select a bin size equal to approximately the width of an expected spectral feature (R=10), using {\tt PandExo} \citep{Batalha2017PANDEXO}. Given the mean precision of our combined Visit 1+2 spectrum is $\overline{\Delta y} = 82$ ppm, and estimating that atmospheric features span 4 scale heights for temperate sub-Neptunes \citep{Brande2024CloudsExoplanets}, we find a lower limit of $\mu>3.9\pm0.8$\,u. This lower limit on $\mu$ is an improvement over the equivalent limit we can calculate from the HST data \citep{deWit2018} of $\mu>2.59\pm0.05$\,u. Here, we have taken $M_p = 0.772 M_{\oplus}$ (where $M_{\oplus}=5.972\times10^{24}$\,kg), $R_p=0.910 R_{\oplus}$ (where $R_{\oplus}=6.3781\times10^{6}$\,m), $R_{*}=0.117 R_{\astrosun}$ (where $R_{\astrosun}=6.957\times10^{8}$\,m), and $T = 249.7$\,K \citep{Ducrot2020}. 

Stronger limits result from considering only the data larger than 1~\,$\micron$, where less stellar contamination is expected \citep{Seager2024WhySpectroscopy}. Visits 1+2 then have a reduced mean uncertainty of 53\,ppm, resulting in $\mu>6.0\pm0.2$\,u, which is stronger than limits derived from HST \citep{deWit2018, Moran2018}. For completeness, we further calculate limits on $\mu$ for $n = $2--5 scale heights, as summarized in Table~\ref{tab:mu_table}. Finally, using the combined four-transit, stellar-contamination-mitigated spectrum from \cite{Espinoza2025}, we find our strongest lower limit of $\mu>8.6\pm0.4$\,u. An atmosphere at the lower end of this limit would be rich in lighter gases such as H$_2$ and He, which is unlikely for a terrestrial planet. While these model-agnostic limits are informative, they do not account for spectral features from specific expected atmospheric constituents (e.g., CO$_2$, CH$_4$) nor do they account for the possible presence of clouds and/or hazes. Additionally, they are dependent on the choice of binning. We proceed to consider more realistic spectral models.

\begin{deluxetable*}{lcccccc}
    \renewcommand{\arraystretch}{1.1}
    \tabletypesize{\footnotesize}
    \tablecolumns{5} 
    \tablecaption{\textbf{Model-agnostic limits on TRAPPIST-1~e's atmospheric mean molecular weight}}
    \tablehead{Data & $\overline{\Delta y}$ (ppm) & Uncertainty on $\overline{\Delta y}$ (ppm) & \multicolumn{4}{c}{Lower limit on $\mu$ (u)}} 
    \startdata
    & & & 2H & 3H & 4H & 5H \\
    \hline
    \hline
    HST & 123.14 & 2.21 & $1.29\pm0.02$ & $1.94\pm0.03$ & $2.59\pm0.05$ & $3.23\pm0.06$ \\
    JWST Visits 1+2 & 81.92 & 17.03 & $1.9\pm0.4$ & $2.9\pm0.6$ & $3.9\pm0.8$ & $4.9\pm1.0$ \\
    HST + JWST Visits 1+2 & 86.5 & 15.29 & $1.8\pm0.3$ & $2.8\pm0.5$ & $3.7\pm0.7$ & $4.6\pm0.8$ \\
    JWST Visits 1+2, $\textgreater 1 \micron$ & 53.22 & 1.56 & $3.0\pm0.1$ & $4.5\pm0.1$ & $6.0\pm0.2$ & $7.5\pm0.2$ \\ 
    HST + JWST Visits 1+2, $\textgreater 1 \micron$ & 62.86 & 4.69 & $2.5\pm0.2$ & $3.8\pm0.3$ & $5.1\pm0.4$ & $6.3\pm0.5$ \\ 
    JWST Visits 1--4, Decontaminated & 59.33 & 11.59 & $2.7\pm0.5$ & $4.0\pm0.8$ & $5.4\pm1.0$ & $6.7\pm1.3$ \\
    JWST Visits 1--4, Decontaminated, $\textgreater 1 \micron$ & 36.96 & 1.85 & $4.3\pm0.2$ & $6.5\pm0.3$ & $8.6\pm0.4$ & $10.8\pm0.5$ \\[2pt]
    \enddata
    \label{tab:mu_table}
    \tablecomments{The lower limit on $\mu$ is calculated for the HST and JWST visits 1 and 2 in various combinations following the equation in Section \ref{sec:mu_results}. We calculate $\mu$ for 2 to 5 scale heights (H). To match expected feature sizes, we bin to R=10.}
\end{deluxetable*}

\begin{figure*}
    \centering
    \includegraphics[width=\linewidth]{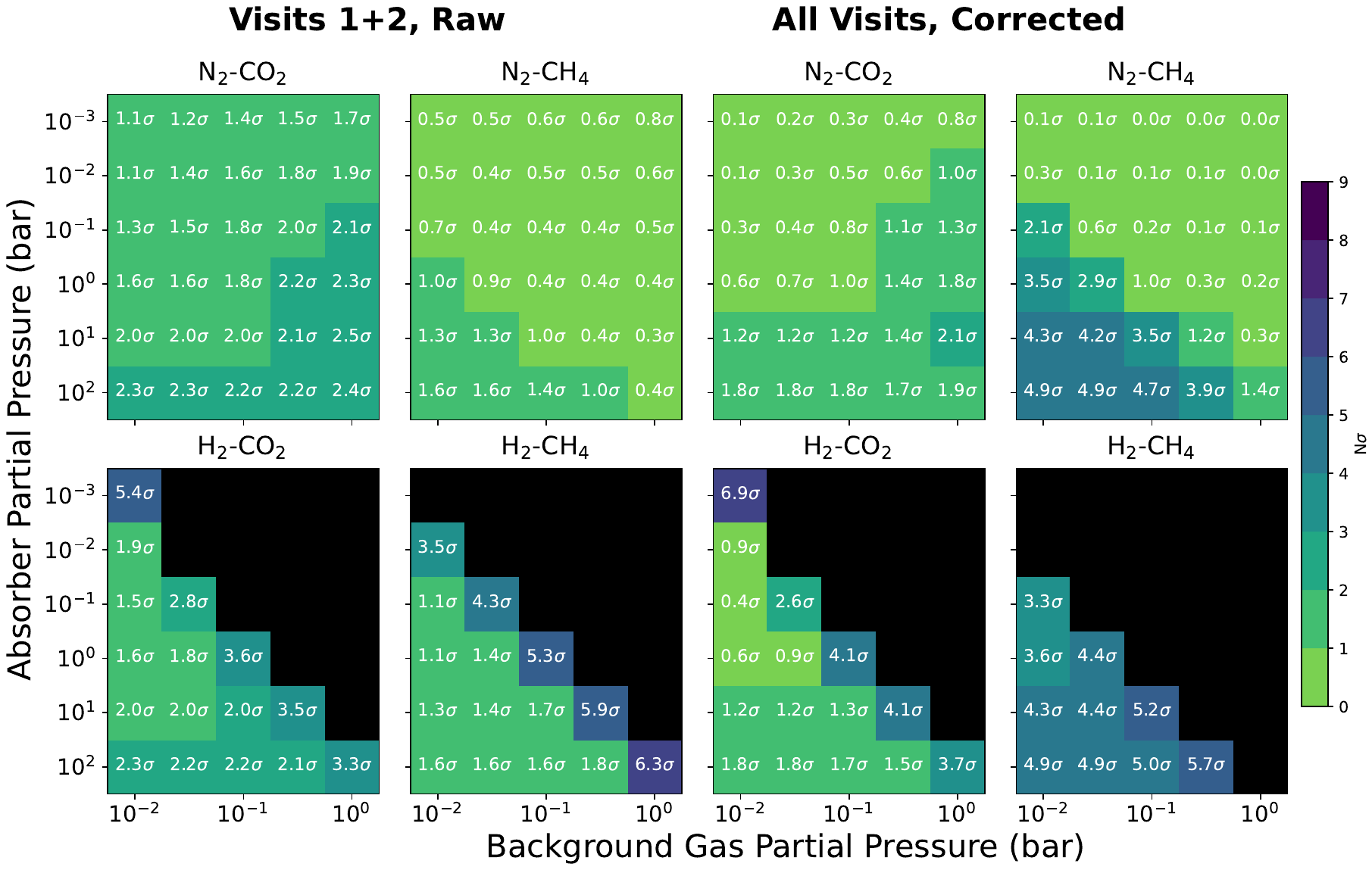}
    \caption{\textbf{Rejection significance for atmospheric forward models compared to TRAPPIST-1~e's JWST transmission spectra.} Each subplot represents a background gas (N$_2$ or H$_2$) together with an absorber (CO$_2$ or CH$_4$) over a range of surface partial pressures shown on the x and y-axes. Gases are shown above each subplot. Grid boxes are labeled and colored with the significance of the difference between the data and the forward model. Black-colored boxes represent ``infinite"$\sigma$, meaning that the models are firmly inconsistent with the data and can be ruled out. The four boxes on the left side of the figure are for the combined Visits 1 and 2, which were naturally less impacted by stellar contamination, while the four boxes on the right side are for the GP stellar-contamination-corrected spectrum from \cite{Espinoza2025}, which includes all four visits. In both cases, our data are consistent across a range of N$_2$ atmospheres, but we are able to place additional constraints on H$_2$-rich atmospheres. In particular, we can rule out H$_2$-rich atmospheres with a strong absorber until increasing the amount of the heavier absorber flattens out the spectrum so that any possible features are buried in the uncertainty. When all four transits are combined and stellar contamination is (partially) mitigated, we are able to place moderately tighter constraints on atmospheres with CH$_4$ than we could with just Visits 1 and 2 combined.}
    \label{fig:forwardmodelingsummary}
\end{figure*}

\subsection{Atmospheric Constraints from \\ Clear-sky Forward Models} \label{sec:forward_model_grid}

We next consider the range of possible atmospheres on TRAPPIST-1~e from a detailed comparison of the data with grids of forward modeled transmission spectra. Our approach is similar to attempts to interpret secondary eclipse photometric emission measurements of the interior TRAPPIST-1 planets (e.g., \citealt{Ih2023}). 

We calculate the synthetic spectra using {\tt petitRADTRANS} ({\tt pRT}; \citealt{Molliere2019prt}), assuming well-mixed atmospheres with varying levels of CO$_2$, CH$_4$, H$_2$, and N$_2$. We employ simple temperature-pressure profiles corresponding to dry adiabatic profiles transitioning into an isothermal stratosphere \citep{Pierrehumbert2010}. Our temperature-pressure profiles are an acceptable simplification because transit spectra are not very sensitive to the temperature-pressure structure of the atmosphere, particularly given the low precision of our data. Similarly, we neglect potential trace gases caused by photochemistry, which would be undetectable with our precision as our current goal is to constrain the presence of major atmospheric absorbers and the bulk gas composition. A possible exception to this assumption would be planets in which photochemistry leads to trace gases (e.g., CO, CH$_4$, C$_5$H$_8$) becoming major atmospheric constituents due to photochemical runaway \citep{Kasting1990, Segura2005, Zhan2021, Ranjan2022}.

We model atmospheres with partial pressures for H$_2$, N$_2$, CO$_2$, O$_2$, and CH$_4$ ranging from 0 to 100\,bar, as measured at the surface. The partial pressure of a gas is the product of its volume mixing ratio and the total atmospheric pressure. We include CO$_2$ as it is expected to be common in rocky planet atmospheres due to volcanic outgassing \citep{Gaillard2014, Wogan2020} and is also photochemically stable and expected to be well mixed to high altitudes in planetary atmospheres \citep{Harman2018, Lincowski2018, Ranjan2023}. Further, the strong 4.3\,$\micron$ CO$_2$ spectral feature is detectable for both modern and Archean Earth-like atmospheric compositions \citep{Krissansen-Totton2018, Mikal-Evans2022, Meadows2023}. We include CH$_4$ due to its importance in the spectra of Earth \citep[e.g.][]{Sagan1993} and Titan \citep[e.g.][]{Robinson2014}, alongside its predicted detectability in TRAPPIST-1~e transmission spectra \citep[e.g.,][]{Krissansen-Totton2018,Lustig-Yaeger2019}. We include O$_2$ due to the possibility of abiotic O$_2$ from atmospheric escape on M dwarf planets \citep{Ramirez2014TheStars, Luger2015ExtremeDwarfs}. H$_2$ and N$_2$ are adopted as potential background gases which, while unlikely to generate detectable spectral features in isolation, affect spectra by altering the $\mu$ of the atmosphere and via collision-induced absorption (CIA). 

Figure~\ref{fig:forwardmodelingsummary} summarizes how well our atmospheric scenarios agree with TRAPPIST-1~e's JWST NIRSpec PRISM observations. The figure shows the absorber partial pressure on the y-axis and the background gas partial pressure in bars on the x-axis. The first four subfigures on the left are for the uncorrected Visits 1+2 while the four on the left are for all four visits combined and decontaminated with the GP. Within each colored box is the rejection significance. First, for Visits 1 and 2 with background N$_2$, all of the models are, strictly speaking, in agreement with the data and cannot be ruled out. However, our observations do slightly disfavor high CO$_2$ partial pressures (rejected at $\sim2.5\,\sigma$), but are consistent with lower CO$_2$ partial pressures. Pure CH$_4$ atmospheres (absorber partial pressure of 100\,bars) are more consistent with the observations than pure CO$_2$ atmospheres. Replacing the background gas with H$_2$ results in a notably different picture. Even a small amount of H$_2$ ($p\rm{H_2}\gtrsim 10^{-2}$\,bars) boosts the absorption feature strength of even a minute amount of CO$_2$ ($p\rm{CO_2}\sim 10^{-3}$\,bars), meaning that we should have been able to see evidence for such an atmosphere. Thus, large swaths of CO$_2$--H$_2$ parameter space can be ruled out. The only allowed regions of CO$_2$--H$_2$ parameter space possess a sufficiently high $\mu$ to suppress the 4.3\,$\micron$ CO$_2$ feature, allowing us to rule out models with $\mu \leq 5.82$ at 5$\sigma$ or greater. Similarly, trace CH$_4$ with a H$_2$ background exhibits a large exclusion zone with $\mu \leq 9.0$ at 3$\sigma$ or greater. The slight differences between H$_2$--CO$_2$ and H$_2$--CH$_4$ are due to the lighter mass of CH$_4$ (16\,u) compared with CO$_2$ (44\,u) and because CH$_4$ exhibits weaker absorption features. Additional statistics from our forward model analysis are provided in Appendices~\ref{sec:appendixB} and \ref{sec:appendixC}.

When all four Visits are combined, with stellar contamination mitigated, we obtain stronger constraints on potential atmospheres on TRAPPIST-1~e. While the CO$_2$ rejection significances are nominally similar between the Visits 1+2 and Visits 1--4 cases, the additional transits strengthen the constraints on CH$_4$. We find that thick, pure, clear CH$_4$ atmospheres ($p\rm{CH_4} = 100$\,bar) are ruled out at 4.9$\sigma$, but models with lower abundances of CH$_4$ with higher N$_2$ partial pressures (top right of the top right panel in Figure~\ref{fig:forwardmodelingsummary}) provide the overall best fit from our model grid. For the CO$_2$-H$_2$ cases, we are able to rule out atmospheres with $\mu \leq 5.82$ now at 6.9$\sigma$ or greater. For the CH$_4$-H$_2$ cases considered, all are ruled out at 3$\sigma$ or greater, while those with $\mu \leq 9.0$ are ruled out at 5$\sigma$ or greater. These $\mu$ constraints are in good agreement with our model-agnostic constraints from section~\ref{sec:mu_results} ($\mu \textgreater 8.6\pm0.4$\,u for the corrected visits 1--4 above 1$\micron$) 

\begin{figure*}[!ht]
    \centering
    \includegraphics[width=0.92\linewidth]{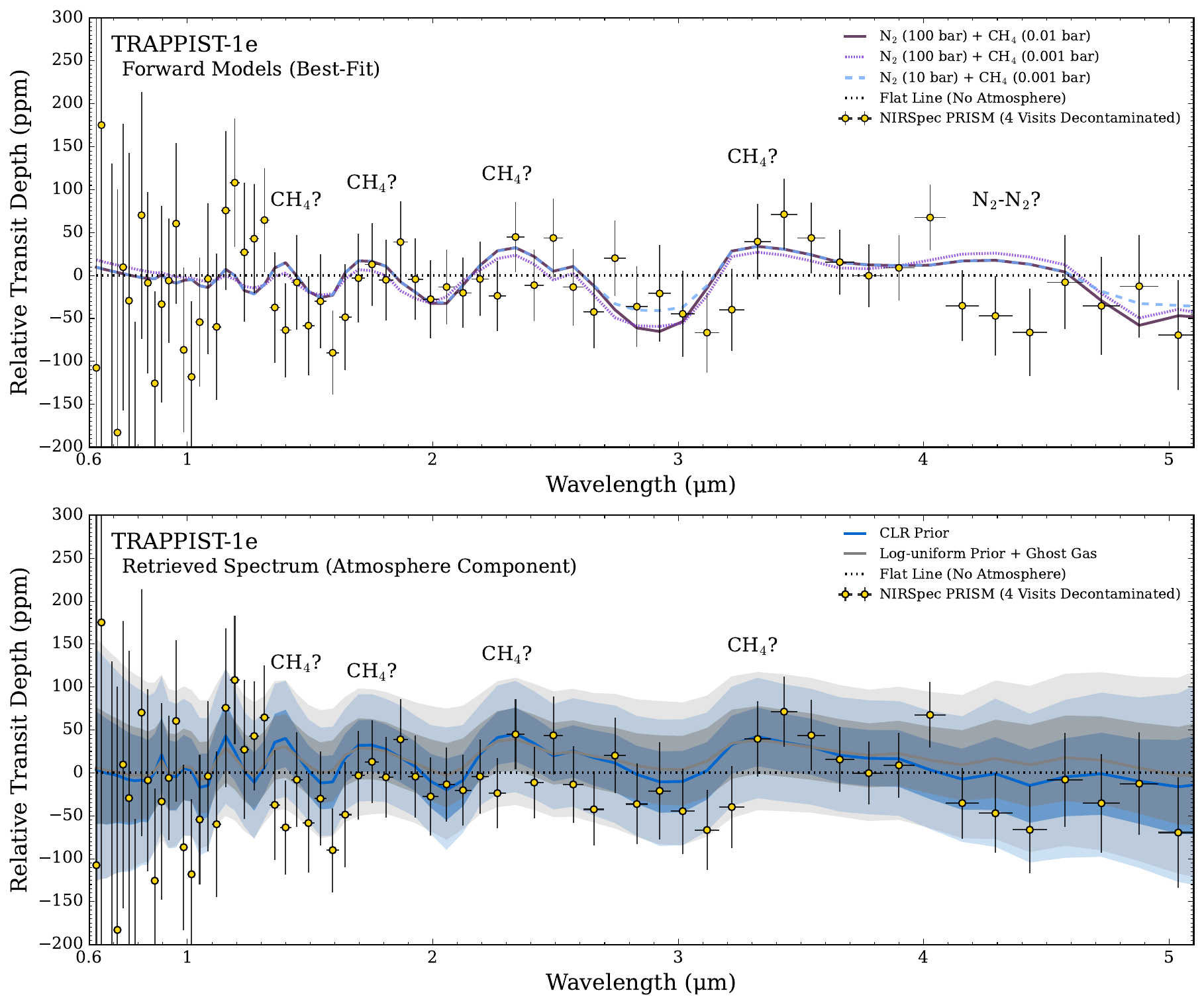}
    \caption{\textbf{Spectral fits to TRAPPIST-1~e's stellar-contamination-corrected transmission spectrum}. Top: best-fitting forward models for three different partial pressures of N$_2$ and CH$_4$ (solid, dotted, and dashed colored lines) compared to a flat line (dotted black line). Bottom: GP+atmosphere retrievals for the centered log-ratio prior (CLR; blue) and log-uniform priors with a `ghost' background gas (gray) compared to a flat line (dotted black line). All models are plotted binned to the same spectral resolution as the data. The wavelengths of potential CH$_4$ absorption bands are annotated. The corresponding corner plot is in Appendix \ref{sec:appendixE}. The best-fitting forward models and both retrieval approaches independently identify spectral features tentatively attributed to CH$_4$ features in a potentially N$_2$-rich atmosphere.}
    \label{fig:spec-fit}
\end{figure*}

We further investigate our best-fitting forward model in Figure~\ref{fig:spec-fit}. Trace quantities of CH$_4$ in a N$_2$-rich atmosphere provide an acceptable fit to TRAPPIST-1~e's Visits 1--4 GP-corrected transmission spectrum. While intriguing to interpret as a hint of CH$_4$, we stress that the present data cannot reject the no-atmosphere null hypothesis \citep{Espinoza2025}. Future observations (JWST Program GO~6456+9256) will reveal if these spectral modulations are instrumental, stellar, or atmospheric. 

\section{Retrieval Analysis: Joint Stellar Contamination--Exoplanet Atmosphere Inferences} \label{sec:jointgpretrievals} 

Next in our hierarchy of interpretation complexity are atmospheric retrievals, which use a Bayesian framework to infer constraints on planetary parameters. One of the key lessons from JWST transmission spectroscopy of the TRAPPIST-1 system (\cite{Espinoza2025}; \citealt{Lim2023, Radica2025}) is that stellar contamination is the dominant feature in transmission spectra of the TRAPPIST-1 planets. Therefore, stellar contamination must be accounted for when exploring possible atmospheric compositions of TRAPPIST-1~e. \cite{Espinoza2025} presented a methodology using GPs to enable simultaneous constraints on both the visit-to-visit stellar contamination signal and any constant-in-time planetary atmosphere signal. Here, we present the constraints obtained from that retrieval framework in detail, showcasing the range of possible atmospheres on TRAPPIST-1~e. 

\subsection{Model Configuration} \label{sec:retrieval-config}

The main retrieval configuration we use, along with the prior distributions we employ for each parameter, is introduced and explained in detail in \cite{Espinoza2025}. We here summarize the general aspects of the framework, along with modifications we did to test the reliability of some of the constraints we put on TRAPPIST-1~e's atmosphere in this work. 

The retrievals use the POSEIDON library \citep{poseidon1, poseidon2} as a forward model for our terrestrial exoplanet atmospheres and a GP framework via the \texttt{george} library \citep{george} using a Mat\`ern 3/2 kernel to account for stellar contamination in each of the spectra we obtain from our four visits. We fit an offset along with the GP hyperparameters and atmospheric constraints to account for the fact that we might lose absolute transit depth information due to stellar contamination in our observations. We also fit for an extra ``jitter" term on each of our visits, which accounts for underestimated error bars in the spectrum in each of our visits. In this way, the framework takes into account the uncertain nature of stellar contamination. We correct for two types of contamination: spectral shape through the GP and absolute transit depths through the visit-dependent offsets. For the exoplanet atmospheric model, we assume an isothermal atmosphere where the atmospheric temperature has a uniform prior, and we fit for both a reference pressure and an effective surface pressure (i.e., a surface or equivalently a cloud-top pressure), where both have log-uniform priors. We use the centered-log ratio (CLR) transformation to model the mixing ratios following \cite{BennekeSeager2013}, where we include H$_2$, CO$_2$, CH$_4$, H$_2$O, N$_2$, O$_2$, O$_3$, N$_2$O and CO \citep{Lin2021, Lustig-Yaeger2023}, setting wide uniform priors on the centered log-ratio parameters. Any of these species can be the background gas. While the presence of condensable species could, in principle, alter the lapse rate through latent heat release, the limited signal-to-noise of our data justifies the use of a simplified isothermal profile.

In addition to the CLR framework, we also employed a ``ghost" background gas mean molecular weight retrieval framework to perform inferences on the data. The ghost gas is a background gas with no spectroscopic features, but a freely fit mean molecular weight, which acts to ``fill" the mean molecular weight of the atmosphere. The motivation for this experiment is to highlight the \textit{spectroscopic} contribution of each of the gases in our retrieval, leaving the task of filling the atmosphere---and thus defining the scale height of the atmosphere---to this ghost background gas. We parameterize this background ghost gas via its molecular weight $\mu_\textrm{ghost}$, for which we define a uniform prior between 2--100\,u, allowing for a broad range of atmospheric compositions from H$_2$-dominated to heavy sulfur compounds such as SO$_2$.  

As the ghost background gas assumes the responsibility of a molecule dominating the atmosphere, we ascribe log-uniform priors on the mixing ratios of our specified molecules instead of CLR priors. As such, we set wide, log-uniform priors between $10^{-12}$ and 1 on the mixing ratios for our spectrally active molecules in this second retrieval configuration. This atmospheric configuration is embedded in the same GP framework defined above. Further details about the retrieval modeling framework can be found in Appendix \ref{sec:appendixD}.

\subsection{TRAPPIST-1~e Compared to Solar System Terrestrial Atmospheres} \label{sec:sscomparison} 

Our retrieval analysis suggests that TRAPPIST-1~e is unlikely to possess a CO$_2$-dominated atmosphere analogous to Venus or Mars based on both abundance and pressure. While our results are consistent with a wide range of atmospheric scenarios---including no atmosphere---our posterior distribution reveals joint effective surface pressure--volume mixing ratio constraints for each molecule that elucidate the types of atmosphere that are best supported by the TRAPPIST-1~e's JWST transmission spectra. While we showed that a flat-line model cannot be rejected for the GP-corrected data (see Section~\ref{Sec:flat_line}), our retrieval analysis can nevertheless place constraints on the abundances of key gases that absorb strongly over the 0.6--5.2\,$\micron$ range covered by the NIRSpec PRISM (in particular, CO$_2$ and CH$_4$).

\begin{figure*}
    \centering
    \includegraphics[width=0.75\linewidth]{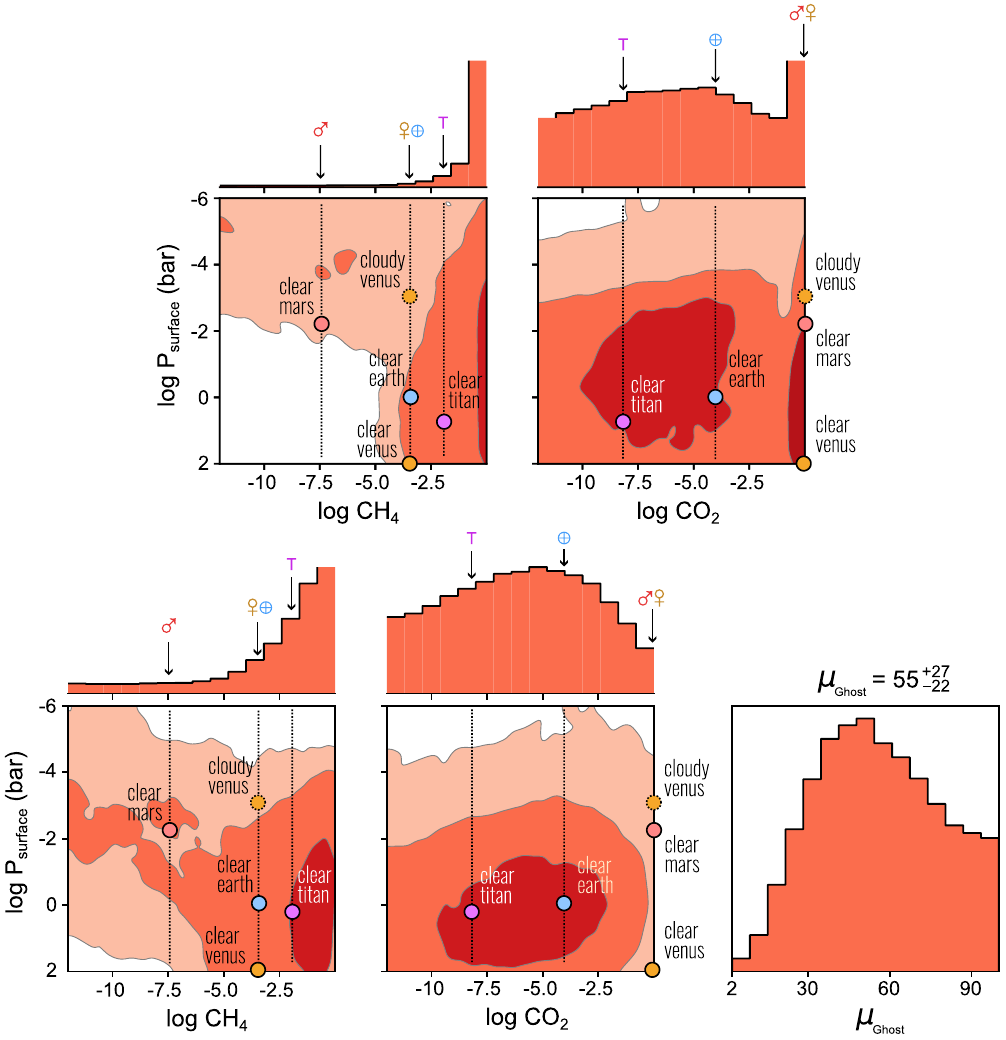}
    \caption{\textbf{\underline{Upper}: posterior for the centered log-ratio (CLR) prior retrievals where every gas is allowed to dominate and \underline{Lower}: agnostic background (ghost) gas retrieval results with log-uniform priors.} The plots show the volume mixing ratio (VMR) for CH$_4$ (first) and CO$_2$ (second) compared with the effective surface pressure (i.e., surface or cloud-top pressure). Contours are drawn and shaded in red for 1, 2, and 3$\sigma$. Above each plot is a histogram of the VMR for each gas. Additionally, for the lower plot, a histogram of the mean molecular weight of the background ghost gas is shown ($\mu_{ghost}$) on the far right. From the CLR retrieval, the CH$_4$ abundances and pressures comparable to Venus, Earth, Mars, and Titan are all allowed, though the clear Venus, Earth, and Titan cases are all favored. The retrieval with the ghost gas confirms this result and further favors the Titan-like scenario at 3$\sigma$. For CO$_2$, a bimodal distribution is present for the CLR retrieval where a 100\% CO$_2$ atmosphere is allowed for cold temperatures, which can minimize spectral features. The solar system planets are all consistent with this result at 2 to 3$\sigma$. When the ghost background gas is included, the bimodal solution is removed, allowing a weak rejection of the clear Venus and Mars and cloudy Venus at 2$\sigma$.}
    \label{fig:retrieval}
\end{figure*}

First, we consider the results of the GP retrieval with CLR priors. In Figure \ref{fig:retrieval}, we show the posterior distributions for the $\log_{10}$ volume mixing ratio (VMR) of CH$_4$ (left) and CO$_2$ (right) relative to the effective surface pressure. Above each plot is a histogram of the VMRs marginalized over the effective surface pressure parameter. We see that both a 100\% CH$_4$ and a 100\% CO$_2$ atmosphere are consistent with the data when marginalized over all surface pressures. Higher abundances of CH$_4$ are favored across a broad range of temperatures and pressures, while a bimodal distribution emerges for CO$_2$ in pressure space. Up to a 100\% CO$_2$ scenario is allowed, but only for low temperatures where any spectral feature would be minimized. Additionally, the temperature range considered also includes temperatures where CO$_2$ would condense out of the atmosphere. Of the spectrally active gases, only CH$_4$ has no upper bound across all pressures. In particular, the posterior for CH$_4$ favors a high abundance, without a strong constraint on the corresponding surface pressure, but the lack of a well-defined lower limit on the CH$_4$ abundance indicates that this molecule is not detected with the present data. We can consider how these constraints compare to solar system atmospheres.

Regions of parameter space relevant to solar system atmospheres are disfavored when considering the dependence on the effective surface pressure. Clear Venus, Earth, and Titan-like abundances and pressures for CH$_4$ lie within the 2--3$\sigma$ credible region while clear Mars and cloudy Venus are in the 1--2$\sigma$ credible region. There is no CH$_4$-rich solar system terrestrial planet to serve as an analog, but Titan is the most comparable with $\sim6$\% CH$_4$ \citep{Niemann2010}. However, TRAPPIST-1~e is significantly warmer than Titan and receives more stellar irradiation. Furthermore, our CO$_2$ constraints allow for clear Venus, Earth, Mars, and Titan as well as cloudy Venus. However, the cases that allow for large amounts of CO$_2$ in the atmosphere correspond with temperatures where CO$_2$ would freeze out, rendering them implausible. In Appendix \ref{sec:appendixF}, Figure \ref{fig:phasediagram}, we show the pressure-temperature posterior together with the phase diagram for CO$_2$.

Second, we consider the GP retrieval with the ghost background gas. In Figure~\ref{fig:retrieval}, we see that the posterior for CH$_4$ similarly exhibits a stronger preference for a high CH$_4$ abundance. The inclusion of the ghost background gas demonstrates that the CH$_4$ peak seen in Figure~\ref{fig:retrieval} is not caused by the CLR prior, but rather there are spectroscopic features in TRAPPIST-1~e's transmission spectra that are consistent with CH$_4$ absorption (see Figure~\ref{fig:spec-fit}). We explore the implications of the potential presence of CH$_4$ further in Section~\ref{sec:CH4?}. Using a ghost background gas, we now find somewhat stricter upper limit on the CO$_2$ abundance, including the removal of the 3$\sigma$ confidence interval consistent with a 100\% CO$_2$ scenario. We now find that a clear Mars and Venus and cloudy Venus lie outside the 2$\sigma$ confidence interval and can be weakly rejected. Furthermore, we show that the histogram for $\mu_{\text{ghost}}$ peaks at $55^{+27}_{-22}$\,u, which could suggest one or more non-detected ghost molecules that are heavier than CH$_4$ (16\,u) with a significant abundance. Our findings for other gases are discussed in Appendix~\ref{sec:otheratms}. As shown in Table~\ref{tab:logz}, we compare the log evidences between the three retrieval models. Confirming the results of \cite{Espinoza2025}, the flat-line retrievals are preferred over the CLR ($\Delta \log \text{Z} = 5.3$). However, the flat line is only marginally preferred over the ghost background gas scenario ($\Delta \log \text{Z} = 0.5$).

If an atmosphere exists on TRAPPIST-1~e, our data suggest that it is best matched by a relatively heavy, spectrally inactive gas together with CH$_4$. However, we stress that the evidence neither warrants the detection of an atmosphere nor rules one out. In Figure~\ref{fig:spec-fit}, we show that both the CLR and ghost gas retrievals identify CH$_4$ bands as candidate spectral features. Our retrievals invoke these features to explain a time-independent spectral component, consistent across the four TRAPPIST-1~e transits, that are not fit by the visit-by-visit GP. This N$_2$--CH$_4$ scenario is consistent with the best-fit models independently identified by our forward modeling analysis. Nevertheless, while intriguing, we emphasize that additional JWST observations will be required to unambiguously detect an atmosphere on TRAPPIST-1~e.

\begin{deluxetable}{lccc}
    \renewcommand{\arraystretch}{0.85}
    \tabletypesize{\footnotesize}
    \tablecolumns{3} 
    \tablecaption{\textbf{Retrieval Model Comparison}}
    \tablehead{Model & $\log \text{Z}$ & $\Delta \log \text{Z}$ & $\sigma$-significance} 
    \startdata
    \hline
    \hline
    Flat line & 533.93 & --- & --- \\
    CLR & 528.61 & 5.3 & 3.7 \\
    Log-uniform + Ghost $\mu$& 533.42 & 0.5 & 1.68 \\
    \enddata
    \tablecomments{Comparison of joint GP retrieval models using the log evidence values.}
    \label{tab:logz}
\end{deluxetable}


\section{Summary and Discussion}\label{sec:summary}

We observed four transits of TRAPPIST-1~e with JWST NIRSpec PRISM, finding significant differences between the spectra derived from each visit. The differences unambiguously tell us that the spectra are dominated by stellar contamination, and any potential atmospheric spectral features must be treated with extreme skepticism. Visits 1 and 2 are less impacted by stellar noise than the later two visits. In our companion Letter \citep{Espinoza2025}, we demonstrated how to mitigate the impact of stellar contamination using a GP. Here, we employ a hierarchy of interpretation techniques to constrain possible atmospheric scenarios on TRAPPIST-1~e. Our primary conclusions are as follows.

\begin{enumerate}
    \item TRAPPIST-1~e's transmission spectrum is adequately modeled by a flat line, which could represent a bare rock, a high mean molecular weight atmosphere, or a thick cloud deck. However, there remain statistically significant underlying undulations in the data that are due to either uncorrected stellar contamination, atmospheric features, or correlated instrumental errors.
    \item The lack of evidence for CO$_2$ at 4.3\,$\micron$ disfavors clear Mars-like and Venus-like and cloudy Venus-like compositions and pressures on TRAPPIST-1~e.
    \item We identify a tentative preference for CH$_4$ absorption in a spectrally quiet background gas (e.g., N$_2$) atmosphere, but this evidence does not constitute a statistically significant detection with the present four transits and is, at best, a hint to be investigated further.
\end{enumerate}

We now analyze the implications of these results.

\subsection{An Airless World or a High-$\mu$ Atmosphere \\ Are Equally Likely}

First, we compared our data to a flat line, representative of either an airless body or a high-altitude cloud deck. Despite our inability to remove all aspects of stellar contamination, our combined four-transit spectrum is adequately fit by a flat line ($\log \text{Z} = 534$), agreeing with expectations that atmospheric features would not be clearly visible with only four transits worth of observations. Additionally, the fit suggests that the GP was at least partially successful at mitigating stellar contamination. We can use our spectrum to place constraints on a possible atmosphere in the absence of an identified detection of a specific molecular feature. Second, using a model-agnostic approach, we found a lower limit of $\mu > 5.95^{+0.21}_{-0.23}$\,u. Third, we also compare our uncorrected (Visits 1+2) and decontaminated spectrum (Visits 1-4) to a grid of forward models. In all cases, we find that our data are consistent with a broad range of N$_2$-dominated atmospheres, even considering the impact of strong absorbers like CO$_2$ and CH$_4$. We are able to rule out atmospheres rich in H$_2$, consistent with prior work \citep{deWit2018}.

\subsection{TRAPPIST-1~e is Unlikely to be a \\ Venus or Mars Analog}\label{ref:CO2upperlimit}

The GP retrieval technique presents a promising method to extract meaningful atmospheric constraints despite the impact of unknown stellar contamination. While our data are statistically consistent with a no-atmosphere scenario, the best-fitting atmospheric scenario does show weak evidence for some spectral features (see below). The lack of evidence of CO$_2$, enables us to place limits on its abundance and the allowable surface pressure, which weakly disfavors a thin, Mars-like CO$_2$ atmosphere and a thick, Venus-like CO$_2$ atmosphere at 2$\sigma$ confidence. Additionally, the temperature constraints place an upper limit of about 200 K at 2$\sigma$, in agreement with a scenario unlike Venus. However, despite our best efforts, there are still residual features in the data that are not explained and likely indicate that stellar contamination was not fully mitigated and/or that there are atmospheric features present in the data that lie below the detection threshold. 

\subsection{Speculative Evidence for Methane} \label{sec:CH4?} 

We have shown that models that include CH$_4$ are viable fits to TRAPPIST-1e's transmission spectrum through both our forward model analysis and retrievals. However, we stress that the statistical evidence falls far below that required for a detection. While an atmosphere containing CH$_4$ and a (relatively) spectrally quiet background gas (e.g., N$_2$) provides a good fit to the data, these initial TRAPPIST-1~e transmission spectra remain consistent with a bare rock or cloudy atmosphere interpretations. Additionally, we note that our ``best-fit" CH$_4$ model does not explain all of the correlated features present in the data. Here we briefly examine the theoretical plausibility of a N$_2$-CH$_4$ atmosphere on TRAPPIST-1~e to contextualize our findings.

If future observations substantiate a N$_2$-CH$_4$ atmosphere on TRAPPIST-1~e, then TRAPPIST-1~e may be analogous to our solar system's Titan, though in a different temperature regime. Titan hosts a N$_2$-dominated, CH$_4$-rich ($1-5\%$; \citealt{Niemann2005}) atmosphere, thought to be sourced geochemically, e.g. via outgassing of natal CH$_4$ \citep{Tobie2006, Nixon2024}. This mechanism may be less tenable for TRAPPIST-1~e due to its higher instellation level and consequent possibility of a much shorter methane lifetime; further work is required to elucidate this possibility \citep{Thompson2022}. Volcanism and serpentinization are other potential source of CH$_4$ \citep{GuzmanMarmolejo2013, Wogan2020}. Additionally, it is likely that a CH$_4$-rich atmosphere would be prone to haze formation, which would mute spectral features and lead to an upward slope towards the near-infrared \citep[NIR; e.g.,][]{Arney2016, Morley2017, Fauchez2019}. UV stellar irradiation photolyzes CH$_4$, forming haze precursors. Hydrocarbon hazes then form at high altitudes. As we do not see a large scattering slope towards shorter wavelengths, we make the simplifying assumption of being haze-free. Ongoing transmission spectrum observations of TRAPPIST-1~e (e.g. GO~6456+9256) will allow this potential presence of CH$_4$ to be either disproven or confirmed. 

\subsection{Limited Constraints on the Habitability of TRAPPIST-1~e}

TRAPPIST-1~e's instellation ($S_p/S_\oplus=0.66\pm0.05$, $S_\oplus\equiv$ Earth solar constant; \citealt{Gillon2017}) allows for the potential presence of a global surface ocean. However, this is contingent on an atmospheric composition capable of supporting the necessary thermal conditions \citep{Kasting1993HabitableStars, Kopparapu2013HabitableEstimates}. As TRAPPIST-1~e receives substantially less instellation compared to Earth, TRAPPIST-1~e requires substantially more greenhouse warming compared to Earth to be globally habitable (surface liquid water stable across most of the planet). Assuming CO$_2$ to be the main greenhouse gas, climate studies find pCO$_2>0.002-0.1$ bar is required for surface liquid water to be stable across $>50\%$ of the planet's surface \citep{Wolf2017, Turbet2018, Sergeev2022} with the range driven by variation in assumed pN$_2$ and methodological differences in the treatment of NIR snow/ice albedo and atmospheric radiative transfer. Our constraints on CO$_2$ allow us to begin to assess the likelihood of an ice-free ocean on TRAPPIST-1~e. 

\begin{figure*}[!ht]
    \centering
    \includegraphics[width=0.8\textwidth]{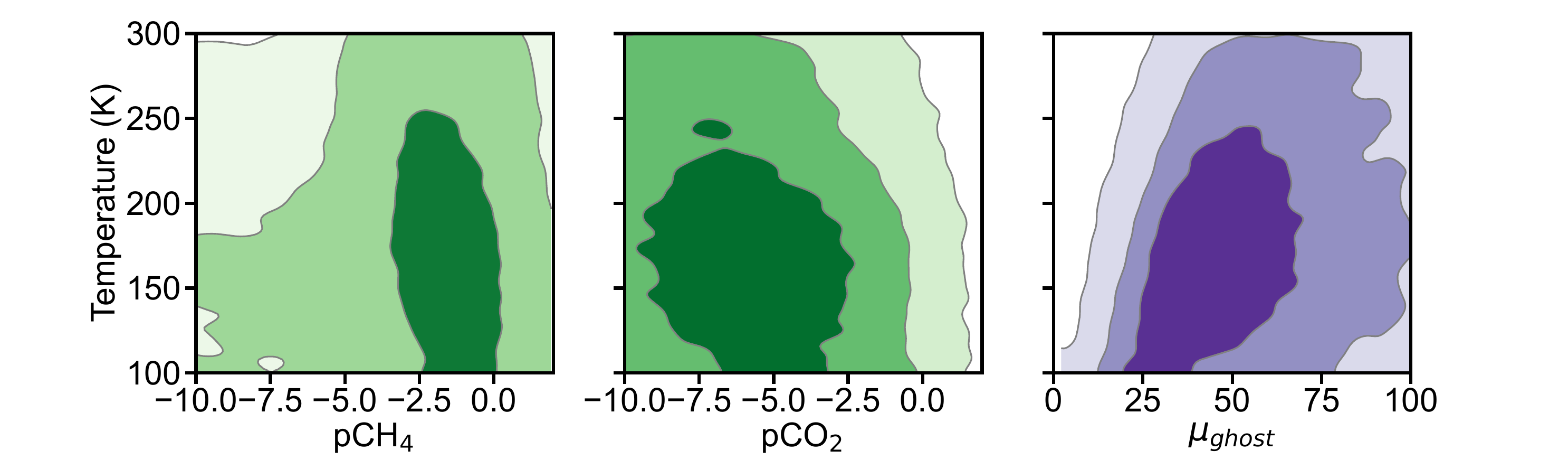}
    \caption{\textbf{Posteriors in temperature space for pCH$_4$ and pCO$_2$ (green) along with $\mu_{ghost}$ from the agnostic ghost background gas retrieval.} Temperatures in kelvins are given on the y-axis and partial pressures of CH$_4$ and CO$_2$ are shown on the x-axis (unlike the earlier posteriors, which showed VMR). In temperature space, an upper limit on pCH$_4$ is not found, while pCO$_2 > 0.7$ bar is ruled out at $>2\sigma$ and pCO$_2 > 44$ bars is ruled out at $>3\sigma$. Although the stellar irradiation received by TRAPPIST-1~e permits the possibility of habitability, the presence of liquid surface water requires an atmospheric composition sufficient to sustain greenhouse warming. Based on GCMs \citep{Wolf2017, Turbet2018, Sergeev2022}, our retrieved upper limit on pCO$_2$ allows for the possibility of an ice-free ocean on TRAPPIST-1~e, though cold scenarios are more favored. }
    \label{fig:tempvspch4co2}
\end{figure*}

We weakly disfavor pCO$_2$ of $\gtrsim 0.7$ bar at $>2\sigma$ (Figure \ref{fig:tempvspch4co2}). While our retrieval is suggestive of the possibility of high CH$_4$ on TRAPPIST-1~e, CH$_4$ is a relatively inefficient greenhouse gas on planets orbiting late M dwarfs due to a strong anti-greenhouse effect driven by robust absorption of incoming NIR radiation \citep{Ramirez2018, Turbet2018}. Large amounts of CH$_4$ in a cool atmosphere will form photochemical hazes, which reflect incoming irradiation \citep{McKay1991}. pCH$_4>1$ bar is required for the planet to be globally habitable \citep{Turbet2018}. While such pCH$_4$ is formally permitted by our retrieval within $2\sigma$, lower abundances are weakly preferred by the retrieval. We therefore conclude that while a globally habitable TRAPPIST-1~e is permitted, our posteriors favor a cold scenario. This possibility has received more limited attention in the literature, which has tended to focus on more Earth-like, globally habitable scenarios (e.g., \citealt{Schwieterman2019, Ranjan2022, Meadows2023}); we advocate for increased consideration of cold-planet scenarios. We note, however, that a consensus prediction of global climate models (GCMs) is the stability of liquid water at the substellar point regardless of atmospheric composition \citep{Turbet2018, Sergeev2022}, meaning that if it has an atmosphere, then TRAPPIST-1~e is likely to be locally habitable at the substellar point even if it is globally uninhabitable.

\subsection{Caveats of Stellar Contamination \\ Correction and the Path Forward}

An outstanding question is how successful the GP was at mitigating the impact of stellar contamination. The GP assumes that the non-time-varying part of the spectra is all from the atmosphere, while in reality, some of that may be attributed to the star. For example, there could be a consistent cool component of the star, and the GP may only be mitigating a stochastic hot region. \citet{Rathcke2025} found large-scale granulation on the surface of TRAPPIST-1 that we likely do not fully correct for. In particular, assessing how the 2.5 to 3.5~\micron~region (where a CH$_4$ feature could exist) varies with increased stellar activity is essential. Additional observations will allow us to probe both the validity of our atmospheric constraints and the robustness of the GP stellar contamination mitigation methods. 

Ongoing observations to collect 15 additional transits are underway as part of JWST Program GO~6456+9256  (PIs: Allen \& Espinoza). This program is observing back-to-back transits of TRAPPIST-1~b and e in close succession. As both planets transit nearly the same chord of the star, and TRAPPIST-1~b emission observations point to it being a bare rock \citep{Greene2023}, we can remove the non-time-varying stellar features by dividing the transmission spectrum of TRAPPIST-1~e by TRAPPIST-1~b without the need for a model of the host star. These observations may help us to confirm our upper limit on CO$_2$ and, potentially, find additional evidence of CH$_4$ and N$_2$.

Additionally, these upcoming JWST observations of TRAPPIST-1~e will allow us to test the dependability of our GP stellar contamination mitigation in two ways. First, we can perform the joint GP retrieval analysis on TRAPPIST-1~b and TRAPPIST-1~e separately to see if any of the same non-time-varying features appear in the spectra of both TRAPPIST-1~b and TRAPPIST-1~e. As TRAPPIST-1~b is likely a bare rock, features present in both planet's ``corrected" spectrum would provide evidence that the features are more likely to be stellar in origin, rather than planetary. To break this degeneracy, we can perform a second test comparing transits separated by different lengths in time to identify stellar variability on different timescales (as we do not yet know the timescale(s) to expect for every kind of stellar variability). We can further perform leave-one-out validation tests and compare the results for our posterior distributions and the mean molecular weight.

Our initial atmospheric reconnaissance of TRAPPIST-1~e highlights the enormous promise of observing this habitable zone world with JWST. With 15 additional JWST transits of TRAPPIST-1~e on the horizon, we stand on the precipice of revealing an atmosphere, or its absence, on one of the most compelling rocky exoplanets. In either case, the answer will profoundly impact how we approach the search for habitable worlds with JWST. 

\section*{Acknowledgments}

We would like to thank the anonymous reviewer for the insightful comments, which greatly improved the manuscript. This Letter reports work carried out in the context of the JWST Telescope Scientist Team (\url{https://www.stsci.edu/~marel/jwsttelsciteam.html}; PI: M. Mountain). Funding is provided to the team by NASA through grant 80NSSC20K0586. Based on observations with the NASA/ESA/CSA JWST, associated with program GTO-1353 (PI: N. K. Lewis), obtained at the Space Telescope Science Institute, which is operated by AURA, Inc., under NASA contract NAS 5-03127. The JWST data presented in this Letter were obtained from the Mikulski Archive for Space Telescopes (MAST) at the Space Telescope Science Institute. The specific observations analyzed can be accessed via  \dataset[DOI: 10.17909/7d90-w905]{http://dx.doi.org/10.17909/7d90-w905}. Retrievals associated with this Letter can be found at \url{https://github.com/nespinoza/TRAPPIST-1~e-GTO-2025}. The authors acknowledge the MIT SuperCloud and Lincoln Laboratory Supercomputing Center for providing HPC resources that have contributed to the research results reported within this Letter. R.J.M. is supported by NASA through the NASA Hubble Fellowship grant HST-HF2-51513.001, awarded by the Space Telescope Science Institute, which is operated by the Association of Universities for Research in Astronomy, Inc., for NASA, under contract NAS 5-26555. K.B.S. performed work as part of the Consortium on Habitability and Atmospheres of M dwarf Planets (CHAMPs) team, supported by the National Aeronautics and Space Administration (NASA) under grant Nos. 80NSSC21K0905 and 80NSSC23K1399 issued through the Interdisciplinary Consortia for Astrobiology Research (ICAR) program. D.R.L acknowledges support from NASA under award number 80GSFC24M0006. C.I.C. acknowledges support by NASA Headquarters through an appointment to the NASA Postdoctoral Program at the Goddard Space Flight Center, administered by ORAU through a contract with NASA. 

\vspace{5mm}

\facilities{JWST (NIRSpec PRISM)}

\software{
\texttt{petitRADTRANS} v2.7.7 \citep{Molliere2019prt} \\
\texttt{PandExo} \citep{Batalha2017PANDEXO}} \\
\texttt{POSEIDON} \citep{poseidon1,poseidon2} \\

\bibliography{t1e_references.bib}{}
\bibliographystyle{aasjournal}

\appendix

\section{Combination of Spectral Data}\label{sec:appendixA}

We use the data from \cite{Espinoza2025}, which have been binned to $R=30$ using the built-in wrapper of {\tt SpectRes} \citep{Carnall2017} within {\tt POSIEDON}. We chose $R=30$ (spectral spacing of $0.04~\micron$) to approximately match the resolution adopted in the earlier HST analysis \citep{deWit2018}. We then recenter each spectrum around zero by subtracting off the error-weighted mean $\bar{y}$, i.e. we calculate $y^{'}_{i}=y_i-\bar{y}$, where $y_{i}$ are the binned R=30 data. We calculated $\bar{y}$ as \citep{BevingtonRobinson2003}: 
\begin{equation}
    \bar{y}=\frac{\sum \frac{y_i}{{\Delta y_i}^{2}}}{\sum \frac{1}{{\Delta y_i}^{2}}}
\end{equation}
\noindent where $\Delta y_{i}$ is the uncertainty on each $y_i$. We then compare the $y^{'}_{i}$ to forward models, as detailed in the remaining sections. We take the uncertainties on $y^{'}_i$ to be the same as those on $y_i$ (i.e. $\Delta y^{'}_i=\Delta y_i$).

Our observations encompass four transits. Visits 1 and 2 are less impacted by stellar activity and are dominated by spots \citep{Espinoza2025}. In contrast, Visits 3 and 4 show higher levels of stellar activity, dominated by faculae. Evidence of significant stellar activity remains in all four transits, but Visits 3 and 4 are particularly impacted. We therefore deem Visits 3 and 4 to be less informative for constraining planetary atmospheric parameters and perform our model comparisons using Visits 1 and 2 alone as a first step before using the GP-corrected, combined Visits 1--4.

We compute a combined Visit 1 ($y^{'}_{1,i}$) and 2 ($y^{'}_{2,i}$) spectrum, $y_{combined, i}$, by calculating the error-weighted mean and adding the errors in quadrature:

\begin{align}
    y_{combined, i}=\frac{y^{'}_{1,i}/{\Delta y_{1,i}}^2 + y^{'}_{2,i}/{\Delta y_{2,i}}^2}{{1/{\Delta y_{1,i}}}^2 + 1/{\Delta y_{2,i}}^2} \\
    \Delta y_{combined, i}=0.5\times\sqrt{{\Delta y_{1,i}}^2 + {\Delta y_{2,i}}^2}
\end{align}
\noindent Here, we have implicitly assumed that our uncertainties are normally distributed \citep{BevingtonRobinson2003}. The combined multi-transit dataset is again centered to zero using the weighted mean to find y$^{'}_{combined}$. We perform our forward model comparison exercise over a range of resolutions and find that it does not meaningfully alter our conclusions.

\section{Comparison of Uncorrected Visits 1 and 2 to Forward Models}\label{sec:appendixB}

Inferences from our forward model analysis for the uncorrected Visits 1 and 2 are shown in Table \ref{tab:longtablemodels}. Columns represent the partial pressures of CO$_2$, N$_2$, pH$_2$, pO$_2$, pCH$_4$; the mean molecular weight ($\mu$); $\chi^{2}$; degrees of freedom ($\nu$); the reduced $\chi^{2}$ ($\chi_{R}^{2}$); p-value; and the N$\sigma$ significance. The results here are presented and discussed in Section \ref{sec:forward_model_grid}. 

\startlongtable
\begin{deluxetable*}{lcccccccccc}
\centering
\tablecaption{Inferences from forward modeling on uncorrected Visits 1 and 2 \label{tab:longtablemodels}} 
\tablehead{pCO$_2$ [bar] & pN$_2$ [bar] & pH$_2$ [bar] & pO$_2$ [bar] & pCH$_4$ [bar] & $\mu$ [u] & $\chi^{2}$ & $\nu$ & $\chi_{R}^{2}$ & p-value & N$\sigma$}
\startdata
    \hline
    \multicolumn{10}{c}{Pure CO$_2$ Atmospheres} \\
    \hline
    0.001 & 0 & 0 & 0 & 0 & 44.0 &  68.99 & 66 & 1.05 & 0.38 & 0.88 \\
    0.01 & 0 & 0 & 0 & 0 & 44.0 &  71.28 & 66 & 1.08 & 0.31 & 1.02 \\
    0.1 & 0 & 0 & 0 & 0 & 44.0 &  75.26 & 66 & 1.14 & 0.2 & 1.27 \\
    1.0 & 0 & 0 & 0 & 0 & 44.0 &  80.18 & 66 & 1.21 & 0.11 & 1.59 \\
    10.0 & 0 & 0 & 0 & 0 & 44.0 &  85.94 & 66 & 1.3 & 0.05 & 1.96 \\
    100.0 & 0 & 0 & 0 & 0 & 44.0 &  90.55 & 66 & 1.37 & 0.02 & 2.25 \\
    \hline
    \multicolumn{10}{c}{Pure O$_2$ Atmospheres} \\
    \hline
    0 & 0 & 0 & 0.001 & 0 & 32.0 &  66.74 & 66 & 1.01 & 0.45 & 0.75 \\
    0 & 0 & 0 & 0.01 & 0 & 32.0 &  66.72 & 66 & 1.01 & 0.45 & 0.75 \\
    0 & 0 & 0 & 0.1 & 0 & 32.0 &  66.58 & 66 & 1.01 & 0.46 & 0.74 \\
    0 & 0 & 0 & 1.0 & 0 & 32.0 &  65.1 & 66 & 0.99 & 0.51 & 0.66 \\
    0 & 0 & 0 & 10.0 & 0 & 32.0 &  62.62 & 66 & 0.95 & 0.6 & 0.53 \\
    0 & 0 & 0 & 100.0 & 0 & 32.0 &  61.2 & 66 & 0.93 & 0.64 & 0.46 \\
    \hline
    \multicolumn{10}{c}{Pure CH$_4$ Atmospheres} \\
    \hline
    0 & 0 & 0 & 0 & 0.001 & 16.0 &  63.89 & 66 & 0.97 & 0.55 & 0.6 \\
    0 & 0 & 0 & 0 & 0.01 & 16.0 &  66.18 & 66 & 1.0 & 0.47 & 0.72 \\
    0 & 0 & 0 & 0 & 0.1 & 16.0 &  68.65 & 66 & 1.04 & 0.39 & 0.86 \\
    0 & 0 & 0 & 0 & 1.0 & 16.0 &  72.09 & 66 & 1.09 & 0.28 & 1.07 \\
    0 & 0 & 0 & 0 & 10.0 & 16.0 &  76.08 & 66 & 1.15 & 0.19 & 1.32 \\
    0 & 0 & 0 & 0 & 100.0 & 16.0 &  80.51 & 66 & 1.22 & 0.11 & 1.61 \\
    \hline
    \multicolumn{10}{c}{CO$_{2}$ and H$_{2}$ Atmospheres} \\
    \hline
    0.001 & 0 & 0.01 & 0 & 0 & 5.82 &  144.93 & 66 & 2.2 & 0.0 & 5.38 \\
    0.01 & 0 & 0.01 & 0 & 0 & 23.0 &  84.73 & 66 & 1.28 & 0.06 & 1.88 \\
    0.1 & 0 & 0.01 & 0 & 0 & 40.18 &  78.17 & 66 & 1.18 & 0.15 & 1.46 \\
    1.0 & 0 & 0.01 & 0 & 0 & 43.58 &  80.49 & 66 & 1.22 & 0.11 & 1.61 \\
    10.0 & 0 & 0.01 & 0 & 0 & 43.96 &  85.96 & 66 & 1.3 & 0.05 & 1.96 \\
    100.0 & 0 & 0.01 & 0 & 0 & 44.0 &  90.55 & 66 & 1.37 & 0.02 & 2.25 \\
    0.001 & 0 & 0.1 & 0 & 0 & 2.42 &  644.49 & 66 & 9.77 & 0.0 & inf{$^\dagger$} \\
    0.01 & 0 & 0.1 & 0 & 0 & 5.82 &  246.81 & 66 & 3.74 & 0.0 & inf \\
    0.1 & 0 & 0.1 & 0 & 0 & 23.0 &  99.63 & 66 & 1.51 & 0.0 & 2.83 \\
    1.0 & 0 & 0.1 & 0 & 0 & 40.18 &  83.66 & 66 & 1.27 & 0.07 & 1.81 \\
    10.0 & 0 & 0.1 & 0 & 0 & 43.58 &  86.03 & 66 & 1.3 & 0.05 & 1.96 \\
    100.0 & 0 & 0.1 & 0 & 0 & 43.96 &  90.46 & 66 & 1.37 & 0.02 & 2.25 \\
    0.001 & 0 & 1.0 & 0 & 0 & 2.04 &  1284.71 & 66 & 19.47 & 0.0 & inf \\
    0.01 & 0 & 1.0 & 0 & 0 & 2.42 &  1204.34 & 66 & 18.25 & 0.0 & inf \\
    0.1 & 0 & 1.0 & 0 & 0 & 5.82 &  361.52 & 66 & 5.48 & 0.0 & inf \\
    1.0 & 0 & 1.0 & 0 & 0 & 23.0 &  112.08 & 66 & 1.7 & 0.0 & 3.58 \\
    10.0 & 0 & 1.0 & 0 & 0 & 40.18 &  87.14 & 66 & 1.32 & 0.04 & 2.04 \\
    100.0 & 0 & 1.0 & 0 & 0 & 43.58 &  89.72 & 66 & 1.36 & 0.03 & 2.2 \\
    0.001 & 0 & 10.0 & 0 & 0 & 2.0 &  1450.87 & 66 & 21.98 & 0.0 & inf \\
    0.01 & 0 & 10.0 & 0 & 0 & 2.04 &  1757.1 & 66 & 26.62 & 0.0 & inf \\
    0.1 & 0 & 10.0 & 0 & 0 & 2.42 &  1580.4 & 66 & 23.95 & 0.0 & inf \\
    1.0 & 0 & 10.0 & 0 & 0 & 5.82 &  365.99 & 66 & 5.55 & 0.0 & inf \\
    10.0 & 0 & 10.0 & 0 & 0 & 23.0 &  110.91 & 66 & 1.68 & 0.0 & 3.51 \\
    100.0 & 0 & 10.0 & 0 & 0 & 40.18 &  88.7 & 66 & 1.34 & 0.03 & 2.14 \\
    0.001 & 0 & 100.0 & 0 & 0 & 2.0 &  1508.01 & 66 & 22.85 & 0.0 & inf \\
    0.01 & 0 & 100.0 & 0 & 0 & 2.0 &  1824.3 & 66 & 27.64 & 0.0 & inf \\
    0.1 & 0 & 100.0 & 0 & 0 & 2.04 &  2204.89 & 66 & 33.41 & 0.0 & inf \\
    1.0 & 0 & 100.0 & 0 & 0 & 2.42 &  1903.55 & 66 & 28.84 & 0.0 & inf \\
    10.0 & 0 & 100.0 & 0 & 0 & 5.82 &  384.48 & 66 & 5.83 & 0.0 & inf \\
    100.0 & 0 & 100.0 & 0 & 0 & 23.0 &  106.64 & 66 & 1.62 & 0.0 & 3.25 \\
    \hline
    \multicolumn{10}{c}{CH$_{4}$ and H$_{2}$ Atmospheres} \\
    \hline
    0 & 0 & 0.01 & 0 & 0.001 & 3.27 &  527.48 & 66 & 7.99 & 0.0 & inf \\
    0 & 0 & 0.01 & 0 & 0.01 & 9.0 &  110.09 & 66 & 1.67 & 0.0 & 3.46 \\
    0 & 0 & 0.01 & 0 & 0.1 & 14.73 &  71.77 & 66 & 1.09 & 0.29 & 1.05 \\
    0 & 0 & 0.01 & 0 & 1.0 & 15.86 &  72.55 & 66 & 1.1 & 0.27 & 1.1 \\
    0 & 0 & 0.01 & 0 & 10.0 & 15.99 &  76.13 & 66 & 1.15 & 0.18 & 1.33 \\
    0 & 0 & 0.01 & 0 & 100.0 & 16.0 &  80.51 & 66 & 1.22 & 0.11 & 1.61 \\
    0 & 0 & 0.1 & 0 & 0.001 & 2.14 &  1437.87 & 66 & 21.79 & 0.0 & inf \\
    0 & 0 & 0.1 & 0 & 0.01 & 3.27 &  829.67 & 66 & 12.57 & 0.0 & inf \\
    0 & 0 & 0.1 & 0 & 0.1 & 9.0 &  124.74 & 66 & 1.89 & 0.0 & 4.3 \\
    0 & 0 & 0.1 & 0 & 1.0 & 14.73 &  76.93 & 66 & 1.17 & 0.17 & 1.38 \\
    0 & 0 & 0.1 & 0 & 10.0 & 15.86 &  76.6 & 66 & 1.16 & 0.18 & 1.36 \\
    0 & 0 & 0.1 & 0 & 100.0 & 15.99 &  80.5 & 66 & 1.22 & 0.11 & 1.61 \\
    0 & 0 & 1.0 & 0 & 0.001 & 2.01 &  1523.69 & 66 & 23.09 & 0.0 & inf \\
    0 & 0 & 1.0 & 0 & 0.01 & 2.14 &  2171.04 & 66 & 32.89 & 0.0 & inf \\
    0 & 0 & 1.0 & 0 & 0.1 & 3.27 &  1086.23 & 66 & 16.46 & 0.0 & inf \\
    0 & 0 & 1.0 & 0 & 1.0 & 9.0 &  143.19 & 66 & 2.17 & 0.0 & 5.29 \\
    0 & 0 & 1.0 & 0 & 10.0 & 14.73 &  81.43 & 66 & 1.23 & 0.1 & 1.67 \\
    0 & 0 & 1.0 & 0 & 100.0 & 15.86 &  80.42 & 66 & 1.22 & 0.11 & 1.6 \\
    0 & 0 & 10.0 & 0 & 0.001 & 2.0 &  1156.45 & 66 & 17.52 & 0.0 & inf \\
    0 & 0 & 10.0 & 0 & 0.01 & 2.01 &  1988.95 & 66 & 30.14 & 0.0 & inf \\
    0 & 0 & 10.0 & 0 & 0.1 & 2.14 &  2847.72 & 66 & 43.15 & 0.0 & inf \\
    0 & 0 & 10.0 & 0 & 1.0 & 3.27 &  1312.48 & 66 & 19.89 & 0.0 & inf \\
    0 & 0 & 10.0 & 0 & 10.0 & 9.0 &  155.31 & 66 & 2.35 & 0.0 & 5.9 \\
    0 & 0 & 10.0 & 0 & 100.0 & 14.73 &  83.68 & 66 & 1.27 & 0.07 & 1.81 \\
    0 & 0 & 100.0 & 0 & 0.001 & 2.0 &  857.38 & 66 & 12.99 & 0.0 & inf \\
    0 & 0 & 100.0 & 0 & 0.01 & 2.0 &  1475.32 & 66 & 22.35 & 0.0 & inf \\
    0 & 0 & 100.0 & 0 & 0.1 & 2.01 &  2580.36 & 66 & 39.1 & 0.0 & inf \\
    0 & 0 & 100.0 & 0 & 1.0 & 2.14 &  3671.05 & 66 & 55.62 & 0.0 & inf \\
    0 & 0 & 100.0 & 0 & 10.0 & 3.27 &  1537.6 & 66 & 23.3 & 0.0 & inf \\
    0 & 0 & 100.0 & 0 & 100.0 & 9.0 &  164.59 & 66 & 2.49 & 0.0 & 6.35 \\
    \hline
    \multicolumn{10}{c}{CO$_{2}$ and N$_{2}$ Atmospheres} \\
    \hline
    0.001 & 0.01 & 0 & 0 & 0 & 29.45 &  71.75 & 66 & 1.09 & 0.29 & 1.05 \\
    0.01 & 0.01 & 0 & 0 & 0 & 36.0 &  73.12 & 66 & 1.11 & 0.26 & 1.14 \\
    0.1 & 0.01 & 0 & 0 & 0 & 42.55 &  75.72 & 66 & 1.15 & 0.19 & 1.3 \\
    1.0 & 0.01 & 0 & 0 & 0 & 43.84 &  80.25 & 66 & 1.22 & 0.11 & 1.59 \\
    10.0 & 0.01 & 0 & 0 & 0 & 43.98 &  85.94 & 66 & 1.3 & 0.05 & 1.96 \\
    100.0 & 0.01 & 0 & 0 & 0 & 44.0 &  90.55 & 66 & 1.37 & 0.02 & 2.25 \\
    0.001 & 0.1 & 0 & 0 & 0 & 28.16 &  74.02 & 66 & 1.12 & 0.23 & 1.19 \\
    0.01 & 0.1 & 0 & 0 & 0 & 29.45 &  76.97 & 66 & 1.17 & 0.17 & 1.38 \\
    0.1 & 0.1 & 0 & 0 & 0 & 36.0 &  78.56 & 66 & 1.19 & 0.14 & 1.48 \\
    1.0 & 0.1 & 0 & 0 & 0 & 42.55 &  80.85 & 66 & 1.23 & 0.1 & 1.63 \\
    10.0 & 0.1 & 0 & 0 & 0 & 43.84 &  86.0 & 66 & 1.3 & 0.05 & 1.96 \\
    100.0 & 0.1 & 0 & 0 & 0 & 43.98 &  90.53 & 66 & 1.37 & 0.02 & 2.25 \\
    0.001 & 1.0 & 0 & 0 & 0 & 28.02 &  77.06 & 66 & 1.17 & 0.17 & 1.39 \\
    0.01 & 1.0 & 0 & 0 & 0 & 28.16 &  80.5 & 66 & 1.22 & 0.11 & 1.61 \\
    0.1 & 1.0 & 0 & 0 & 0 & 29.45 &  83.7 & 66 & 1.27 & 0.07 & 1.81 \\
    1.0 & 1.0 & 0 & 0 & 0 & 36.0 &  84.22 & 66 & 1.28 & 0.06 & 1.85 \\
    10.0 & 1.0 & 0 & 0 & 0 & 42.55 &  85.91 & 66 & 1.3 & 0.05 & 1.96 \\
    100.0 & 1.0 & 0 & 0 & 0 & 43.84 &  90.26 & 66 & 1.37 & 0.03 & 2.24 \\
    0.001 & 10.0 & 0 & 0 & 0 & 28.0 &  79.59 & 66 & 1.21 & 0.12 & 1.55 \\
    0.01 & 10.0 & 0 & 0 & 0 & 28.02 &  82.79 & 66 & 1.25 & 0.08 & 1.76 \\
    0.1 & 10.0 & 0 & 0 & 0 & 28.16 &  86.49 & 66 & 1.31 & 0.05 & 1.99 \\
    1.0 & 10.0 & 0 & 0 & 0 & 29.45 &  89.56 & 66 & 1.36 & 0.03 & 2.19 \\
    10.0 & 10.0 & 0 & 0 & 0 & 36.0 &  88.66 & 66 & 1.34 & 0.03 & 2.13 \\
    100.0 & 10.0 & 0 & 0 & 0 & 42.55 &  89.62 & 66 & 1.36 & 0.03 & 2.19 \\
    0.001 & 100.0 & 0 & 0 & 0 & 28.0 &  81.86 & 66 & 1.24 & 0.09 & 1.69 \\
    0.01 & 100.0 & 0 & 0 & 0 & 28.0 &  84.95 & 66 & 1.29 & 0.06 & 1.89 \\
    0.1 & 100.0 & 0 & 0 & 0 & 28.02 &  88.23 & 66 & 1.34 & 0.04 & 2.11 \\
    1.0 & 100.0 & 0 & 0 & 0 & 28.16 &  91.83 & 66 & 1.39 & 0.02 & 2.34 \\
    10.0 & 100.0 & 0 & 0 & 0 & 29.45 &  94.66 & 66 & 1.43 & 0.01 & 2.51 \\
    100.0 & 100.0 & 0 & 0 & 0 & 36.0 &  92.11 & 66 & 1.4 & 0.02 & 2.35 \\
    \hline
    \multicolumn{10}{c}{CH$_{4}$ and N$_{2}$ Atmospheres} \\
    \hline
    0 & 0.01 & 0 & 0 & 0.001 & 26.91 &  62.36 & 66 & 0.94 & 0.6 & 0.52 \\
    0 & 0.01 & 0 & 0 & 0.01 & 22.0 &  61.25 & 66 & 0.93 & 0.64 & 0.46 \\
    0 & 0.01 & 0 & 0 & 0.1 & 17.09 &  66.1 & 66 & 1.0 & 0.47 & 0.72 \\
    0 & 0.01 & 0 & 0 & 1.0 & 16.12 &  71.66 & 66 & 1.09 & 0.3 & 1.05 \\
    0 & 0.01 & 0 & 0 & 10.0 & 16.01 &  76.0 & 66 & 1.15 & 0.19 & 1.32 \\
    0 & 0.01 & 0 & 0 & 100.0 & 16.0 &  80.47 & 66 & 1.22 & 0.11 & 1.61 \\
    0 & 0.1 & 0 & 0 & 0.001 & 27.88 &  62.87 & 66 & 0.95 & 0.59 & 0.54 \\
    0 & 0.1 & 0 & 0 & 0.01 & 26.91 &  60.91 & 66 & 0.92 & 0.65 & 0.45 \\
    0 & 0.1 & 0 & 0 & 0.1 & 22.0 &  60.45 & 66 & 0.92 & 0.67 & 0.43 \\
    0 & 0.1 & 0 & 0 & 1.0 & 17.09 &  68.58 & 66 & 1.04 & 0.39 & 0.86 \\
    0 & 0.1 & 0 & 0 & 10.0 & 16.12 &  75.31 & 66 & 1.14 & 0.2 & 1.27 \\
    0 & 0.1 & 0 & 0 & 100.0 & 16.01 &  80.12 & 66 & 1.21 & 0.11 & 1.58 \\
    0 & 1.0 & 0 & 0 & 0.001 & 27.99 &  63.62 & 66 & 0.96 & 0.56 & 0.58 \\
    0 & 1.0 & 0 & 0 & 0.01 & 27.88 &  61.52 & 66 & 0.93 & 0.63 & 0.48 \\
    0 & 1.0 & 0 & 0 & 0.1 & 26.91 &  59.71 & 66 & 0.9 & 0.69 & 0.39 \\
    0 & 1.0 & 0 & 0 & 1.0 & 22.0 &  60.62 & 66 & 0.92 & 0.66 & 0.43 \\
    0 & 1.0 & 0 & 0 & 10.0 & 17.09 &  70.38 & 66 & 1.07 & 0.33 & 0.97 \\
    0 & 1.0 & 0 & 0 & 100.0 & 16.12 &  77.87 & 66 & 1.18 & 0.15 & 1.44 \\
    0 & 10.0 & 0 & 0 & 0.001 & 28.0 &  64.84 & 66 & 0.98 & 0.52 & 0.65 \\
    0 & 10.0 & 0 & 0 & 0.01 & 27.99 &  62.6 & 66 & 0.95 & 0.6 & 0.53 \\
    0 & 10.0 & 0 & 0 & 0.1 & 27.88 &  60.68 & 66 & 0.92 & 0.66 & 0.44 \\
    0 & 10.0 & 0 & 0 & 1.0 & 26.91 &  59.05 & 66 & 0.89 & 0.72 & 0.36 \\
    0 & 10.0 & 0 & 0 & 10.0 & 22.0 &  60.26 & 66 & 0.91 & 0.68 & 0.42 \\
    0 & 10.0 & 0 & 0 & 100.0 & 17.09 &  71.11 & 66 & 1.08 & 0.31 & 1.01 \\
    0 & 100.0 & 0 & 0 & 0.001 & 28.0 &  66.99 & 66 & 1.01 & 0.44 & 0.77 \\
    0 & 100.0 & 0 & 0 & 0.01 & 28.0 &  63.89 & 66 & 0.97 & 0.55 & 0.6 \\
    0 & 100.0 & 0 & 0 & 0.1 & 27.99 &  61.68 & 66 & 0.93 & 0.63 & 0.48 \\
    0 & 100.0 & 0 & 0 & 1.0 & 27.88 &  59.87 & 66 & 0.91 & 0.69 & 0.4 \\
    0 & 100.0 & 0 & 0 & 10.0 & 26.91 &  58.43 & 66 & 0.89 & 0.73 & 0.34 \\
    0 & 100.0 & 0 & 0 & 100.0 & 22.0 &  60.17 & 66 & 0.91 & 0.68 & 0.41 \\
    \hline
    \enddata
\tablecomments{Espinoza reduction, Visits 1 and 2 combined, R = 30. $^\dagger$ Models with a N$\sigma$ equal to ``inf" do not agree with the data and are unambiguously ruled out.}
\end{deluxetable*}

\newpage

\section{Comparison of Decontaminated Visits 1 to 4 to Forward Models}\label{sec:appendixC}

Inferences from our forward model analysis for the decontaminated Visits 1-4 are shown in Table \ref{tab:longtablemodels_decontam}. Columns are the same as in Appendix \ref{sec:appendixB}. The data are corrected for stellar contamination as described in Section \ref{sec:retrieval-config}.

\startlongtable
\begin{deluxetable*}{lcccccccccc}
\centering
\tablecaption{Inferences from forward modeling on decontaminated Visits 1--4 \label{tab:longtablemodels_decontam}} 
\tablehead{pCO$_2$ [bar] & pN$_2$ [bar] & pH$_2$ [bar] & pO$_2$ [bar] & pCH$_4$ [bar] & $\mu$ [u] & $\chi^{2}$ & $\nu$ & $\chi_{R}^{2}$ & p-value & N$\sigma$}
\startdata
    \hline
    \multicolumn{10}{c}{Pure CO$_2$ Atmospheres} \\
    \hline
    0.001 & 0 & 0 & 0 & 0 & 44.0 &  46.5 & 66 & 0.7 & 0.97 & 0.04 \\
    0.01 & 0 & 0 & 0 & 0 & 44.0 &  49.44 & 66 & 0.75 & 0.94 & 0.08 \\
    0.1 & 0 & 0 & 0 & 0 & 44.0 &  55.45 & 66 & 0.84 & 0.82 & 0.23 \\
    1.0 & 0 & 0 & 0 & 0 & 44.0 &  63.84 & 66 & 0.97 & 0.55 & 0.59 \\
    10.0 & 0 & 0 & 0 & 0 & 44.0 &  74.72 & 66 & 1.13 & 0.22 & 1.24 \\
    100.0 & 0 & 0 & 0 & 0 & 44.0 &  82.92 & 66 & 1.26 & 0.08 & 1.76 \\
    \hline
    \multicolumn{10}{c}{Pure O$_2$ Atmospheres} \\
    \hline
    0 & 0 & 0 & 0.001 & 0 & 32.0 &  44.05 & 66 & 0.67 & 0.98 & 0.02 \\
    0 & 0 & 0 & 0.01 & 0 & 32.0 &  44.05 & 66 & 0.67 & 0.98 & 0.02 \\
    0 & 0 & 0 & 0.1 & 0 & 32.0 &  44.09 & 66 & 0.67 & 0.98 & 0.02 \\
    0 & 0 & 0 & 1.0 & 0 & 32.0 &  44.23 & 66 & 0.67 & 0.98 & 0.02 \\
    0 & 0 & 0 & 10.0 & 0 & 32.0 &  46.25 & 66 & 0.7 & 0.97 & 0.04 \\
    0 & 0 & 0 & 100.0 & 0 & 32.0 &  51.77 & 66 & 0.78 & 0.9 & 0.13 \\
    \hline
    \multicolumn{10}{c}{Pure CH$_4$ Atmospheres} \\
    \hline
    0 & 0 & 0 & 0 & 0.001 & 16.0 &  61.81 & 66 & 0.94 & 0.62 & 0.49 \\
    0 & 0 & 0 & 0 & 0.01 & 16.0 &  80.43 & 66 & 1.22 & 0.11 & 1.6 \\
    0 & 0 & 0 & 0 & 0.1 & 16.0 &  97.04 & 66 & 1.47 & 0.01 & 2.66 \\
    0 & 0 & 0 & 0 & 1.0 & 16.0 &  111.73 & 66 & 1.69 & 0.0 & 3.56 \\
    0 & 0 & 0 & 0 & 10.0 & 16.0 &  124.48 & 66 & 1.89 & 0.0 & 4.29 \\
    0 & 0 & 0 & 0 & 100.0 & 16.0 &  136.11 & 66 & 2.06 & 0.0 & 4.92 \\
    \hline
    \multicolumn{10}{c}{CO$_{2}$ and H$_{2}$ Atmospheres} \\
    \hline
    0.001 & 0 & 0.01 & 0 & 0 & 5.82 &  176.4 & 66 & 2.67 & 0.0 & 6.9 \\
    0.01 & 0 & 0.01 & 0 & 0 & 23.0 &  69.69 & 66 & 1.06 & 0.35 & 0.93 \\
    0.1 & 0 & 0.01 & 0 & 0 & 40.18 &  60.04 & 66 & 0.91 & 0.68 & 0.41 \\
    1.0 & 0 & 0.01 & 0 & 0 & 43.58 &  64.4 & 66 & 0.98 & 0.53 & 0.62 \\
    10.0 & 0 & 0.01 & 0 & 0 & 43.96 &  74.77 & 66 & 1.13 & 0.22 & 1.24 \\
    100.0 & 0 & 0.01 & 0 & 0 & 44.0 &  82.92 & 66 & 1.26 & 0.08 & 1.76 \\
    0.001 & 0 & 0.1 & 0 & 0 & 2.42 &  1182.29 & 66 & 17.91 & 0.0 & inf{$^\dagger$} \\
    0.01 & 0 & 0.1 & 0 & 0 & 5.82 &  379.02 & 66 & 5.74 & 0.0 & inf \\
    0.1 & 0 & 0.1 & 0 & 0 & 23.0 &  95.89 & 66 & 1.45 & 0.01 & 2.59 \\
    1.0 & 0 & 0.1 & 0 & 0 & 40.18 &  69.92 & 66 & 1.06 & 0.35 & 0.94 \\
    10.0 & 0 & 0.1 & 0 & 0 & 43.58 &  74.68 & 66 & 1.13 & 0.22 & 1.23 \\
    100.0 & 0 & 0.1 & 0 & 0 & 43.96 &  82.73 & 66 & 1.25 & 0.08 & 1.75 \\
    0.001 & 0 & 1.0 & 0 & 0 & 2.04 &  2636.34 & 66 & 39.94 & 0.0 & inf \\
    0.01 & 0 & 1.0 & 0 & 0 & 2.42 &  2412.57 & 66 & 36.55 & 0.0 & inf \\
    0.1 & 0 & 1.0 & 0 & 0 & 5.82 &  631.13 & 66 & 9.56 & 0.0 & inf \\
    1.0 & 0 & 1.0 & 0 & 0 & 23.0 &  121.38 & 66 & 1.84 & 0.0 & 4.11 \\
    10.0 & 0 & 1.0 & 0 & 0 & 40.18 &  75.31 & 66 & 1.14 & 0.2 & 1.27 \\
    100.0 & 0 & 1.0 & 0 & 0 & 43.58 &  81.19 & 66 & 1.23 & 0.1 & 1.65 \\
    0.001 & 0 & 10.0 & 0 & 0 & 2.0 &  3060.12 & 66 & 46.37 & 0.0 & inf \\
    0.01 & 0 & 10.0 & 0 & 0 & 2.04 &  3636.61 & 66 & 55.1 & 0.0 & inf \\
    0.1 & 0 & 10.0 & 0 & 0 & 2.42 &  3236.62 & 66 & 49.04 & 0.0 & inf \\
    1.0 & 0 & 10.0 & 0 & 0 & 5.82 &  637.6 & 66 & 9.66 & 0.0 & inf \\
    10.0 & 0 & 10.0 & 0 & 0 & 23.0 &  120.97 & 66 & 1.83 & 0.0 & 4.09 \\
    100.0 & 0 & 10.0 & 0 & 0 & 40.18 &  79.38 & 66 & 1.2 & 0.12 & 1.54 \\
    0.001 & 0 & 100.0 & 0 & 0 & 2.0 &  3284.54 & 66 & 49.77 & 0.0 & inf \\
    0.01 & 0 & 100.0 & 0 & 0 & 2.0 &  3881.28 & 66 & 58.81 & 0.0 & inf \\
    0.1 & 0 & 100.0 & 0 & 0 & 2.04 &  4601.71 & 66 & 69.72 & 0.0 & inf \\
    1.0 & 0 & 100.0 & 0 & 0 & 2.42 &  3930.72 & 66 & 59.56 & 0.0 & inf \\
    10.0 & 0 & 100.0 & 0 & 0 & 5.82 &  677.3 & 66 & 10.26 & 0.0 & inf \\
    100.0 & 0 & 100.0 & 0 & 0 & 23.0 &  113.68 & 66 & 1.72 & 0.0 & 3.67 \\
    \hline
    \multicolumn{10}{c}{CH$_{4}$ and H$_{2}$ Atmospheres} \\
    \hline
    0 & 0 & 0.01 & 0 & 0.001 & 3.27 &  1206.72 & 66 & 18.28 & 0.0 & inf \\
    0 & 0 & 0.01 & 0 & 0.01 & 9.0 &  212.09 & 66 & 3.21 & 0.0 & inf \\
    0 & 0 & 0.01 & 0 & 0.1 & 14.73 &  108.03 & 66 & 1.64 & 0.0 & 3.34 \\
    0 & 0 & 0.01 & 0 & 1.0 & 15.86 &  113.09 & 66 & 1.71 & 0.0 & 3.64 \\
    0 & 0 & 0.01 & 0 & 10.0 & 15.99 &  124.63 & 66 & 1.89 & 0.0 & 4.29 \\
    0 & 0 & 0.01 & 0 & 100.0 & 16.0 &  136.12 & 66 & 2.06 & 0.0 & 4.92 \\
    0 & 0 & 0.1 & 0 & 0.001 & 2.14 &  3374.1 & 66 & 51.12 & 0.0 & inf \\
    0 & 0 & 0.1 & 0 & 0.01 & 3.27 &  1932.8 & 66 & 29.28 & 0.0 & inf \\
    0 & 0 & 0.1 & 0 & 0.1 & 9.0 &  258.97 & 66 & 3.92 & 0.0 & inf \\
    0 & 0 & 0.1 & 0 & 1.0 & 14.73 &  126.07 & 66 & 1.91 & 0.0 & 4.37 \\
    0 & 0 & 0.1 & 0 & 10.0 & 15.86 &  126.06 & 66 & 1.91 & 0.0 & 4.37 \\
    0 & 0 & 0.1 & 0 & 100.0 & 15.99 &  136.24 & 66 & 2.06 & 0.0 & 4.92 \\
    0 & 0 & 1.0 & 0 & 0.001 & 2.01 &  3659.55 & 66 & 55.45 & 0.0 & inf \\
    0 & 0 & 1.0 & 0 & 0.01 & 2.14 &  5101.05 & 66 & 77.29 & 0.0 & inf \\
    0 & 0 & 1.0 & 0 & 0.1 & 3.27 &  2549.08 & 66 & 38.62 & 0.0 & inf \\
    0 & 0 & 1.0 & 0 & 1.0 & 9.0 &  307.4 & 66 & 4.66 & 0.0 & inf \\
    0 & 0 & 1.0 & 0 & 10.0 & 14.73 &  140.89 & 66 & 2.13 & 0.0 & 5.17 \\
    0 & 0 & 1.0 & 0 & 100.0 & 15.86 &  137.46 & 66 & 2.08 & 0.0 & 4.99 \\
    0 & 0 & 10.0 & 0 & 0.001 & 2.0 &  2857.89 & 66 & 43.3 & 0.0 & inf \\
    0 & 0 & 10.0 & 0 & 0.01 & 2.01 &  4773.56 & 66 & 72.33 & 0.0 & inf \\
    0 & 0 & 10.0 & 0 & 0.1 & 2.14 &  6675.47 & 66 & 101.14 & 0.0 & inf \\
    0 & 0 & 10.0 & 0 & 1.0 & 3.27 &  3088.43 & 66 & 46.79 & 0.0 & inf \\
    0 & 0 & 10.0 & 0 & 10.0 & 9.0 &  346.78 & 66 & 5.25 & 0.0 & inf \\
    0 & 0 & 10.0 & 0 & 100.0 & 14.73 &  152.2 & 66 & 2.31 & 0.0 & 5.74 \\
    0 & 0 & 100.0 & 0 & 0.001 & 2.0 &  2072.37 & 66 & 31.4 & 0.0 & inf \\
    0 & 0 & 100.0 & 0 & 0.01 & 2.0 &  3643.41 & 66 & 55.2 & 0.0 & inf \\
    0 & 0 & 100.0 & 0 & 0.1 & 2.01 &  6172.26 & 66 & 93.52 & 0.0 & inf \\
    0 & 0 & 100.0 & 0 & 1.0 & 2.14 &  8566.15 & 66 & 129.79 & 0.0 & inf \\
    0 & 0 & 100.0 & 0 & 10.0 & 3.27 &  3605.38 & 66 & 54.63 & 0.0 & inf \\
    0 & 0 & 100.0 & 0 & 100.0 & 9.0 &  373.88 & 66 & 5.66 & 0.0 & inf \\
    \hline
    \multicolumn{10}{c}{CO$_{2}$ and N$_{2}$ Atmospheres} \\
    \hline
    0.001 & 0.01 & 0 & 0 & 0 & 29.45 &  50.0 & 66 & 0.76 & 0.93 & 0.09 \\
    0.01 & 0.01 & 0 & 0 & 0 & 36.0 &  52.01 & 66 & 0.79 & 0.9 & 0.13 \\
    0.1 & 0.01 & 0 & 0 & 0 & 42.55 &  56.17 & 66 & 0.85 & 0.8 & 0.25 \\
    1.0 & 0.01 & 0 & 0 & 0 & 43.84 &  63.98 & 66 & 0.97 & 0.55 & 0.6 \\
    10.0 & 0.01 & 0 & 0 & 0 & 43.98 &  74.74 & 66 & 1.13 & 0.22 & 1.24 \\
    100.0 & 0.01 & 0 & 0 & 0 & 44.0 &  82.93 & 66 & 1.26 & 0.08 & 1.76 \\
    0.001 & 0.1 & 0 & 0 & 0 & 28.16 &  53.06 & 66 & 0.8 & 0.88 & 0.16 \\
    0.01 & 0.1 & 0 & 0 & 0 & 29.45 &  57.55 & 66 & 0.87 & 0.76 & 0.3 \\
    0.1 & 0.1 & 0 & 0 & 0 & 36.0 &  60.69 & 66 & 0.92 & 0.66 & 0.44 \\
    1.0 & 0.1 & 0 & 0 & 0 & 42.55 &  65.06 & 66 & 0.99 & 0.51 & 0.66 \\
    10.0 & 0.1 & 0 & 0 & 0 & 43.84 &  74.92 & 66 & 1.14 & 0.21 & 1.25 \\
    100.0 & 0.1 & 0 & 0 & 0 & 43.98 &  82.95 & 66 & 1.26 & 0.08 & 1.77 \\
    0.001 & 1.0 & 0 & 0 & 0 & 28.02 &  56.52 & 66 & 0.86 & 0.79 & 0.27 \\
    0.01 & 1.0 & 0 & 0 & 0 & 28.16 &  62.01 & 66 & 0.94 & 0.62 & 0.5 \\
    0.1 & 1.0 & 0 & 0 & 0 & 29.45 &  67.72 & 66 & 1.03 & 0.42 & 0.81 \\
    1.0 & 1.0 & 0 & 0 & 0 & 36.0 &  70.09 & 66 & 1.06 & 0.34 & 0.95 \\
    10.0 & 1.0 & 0 & 0 & 0 & 42.55 &  74.09 & 66 & 1.12 & 0.23 & 1.2 \\
    100.0 & 1.0 & 0 & 0 & 0 & 43.84 &  82.74 & 66 & 1.25 & 0.08 & 1.75 \\
    0.001 & 10.0 & 0 & 0 & 0 & 28.0 &  59.64 & 66 & 0.9 & 0.7 & 0.39 \\
    0.01 & 10.0 & 0 & 0 & 0 & 28.02 &  64.86 & 66 & 0.98 & 0.52 & 0.65 \\
    0.1 & 10.0 & 0 & 0 & 0 & 28.16 &  71.8 & 66 & 1.09 & 0.29 & 1.05 \\
    1.0 & 10.0 & 0 & 0 & 0 & 29.45 &  77.89 & 66 & 1.18 & 0.15 & 1.44 \\
    10.0 & 10.0 & 0 & 0 & 0 & 36.0 &  77.74 & 66 & 1.18 & 0.15 & 1.43 \\
    100.0 & 10.0 & 0 & 0 & 0 & 42.55 &  81.44 & 66 & 1.23 & 0.1 & 1.67 \\
    0.001 & 100.0 & 0 & 0 & 0 & 28.0 &  66.9 & 66 & 1.01 & 0.45 & 0.76 \\
    0.01 & 100.0 & 0 & 0 & 0 & 28.0 &  70.97 & 66 & 1.08 & 0.32 & 1.0 \\
    0.1 & 100.0 & 0 & 0 & 0 & 28.02 &  76.21 & 66 & 1.15 & 0.18 & 1.33 \\
    1.0 & 100.0 & 0 & 0 & 0 & 28.16 &  83.0 & 66 & 1.26 & 0.08 & 1.77 \\
    10.0 & 100.0 & 0 & 0 & 0 & 29.45 &  88.55 & 66 & 1.34 & 0.03 & 2.13 \\
    100.0 & 100.0 & 0 & 0 & 0 & 36.0 &  84.83 & 66 & 1.29 & 0.06 & 1.89 \\
    \hline
    \multicolumn{10}{c}{CH$_{4}$ and N$_{2}$ Atmospheres} \\
    \hline
    0 & 0.01 & 0 & 0 & 0.001 & 26.91 &  47.62 & 66 & 0.72 & 0.96 & 0.05 \\
    0 & 0.01 & 0 & 0 & 0.01 & 22.0 &  58.58 & 66 & 0.89 & 0.73 & 0.34 \\
    0 & 0.01 & 0 & 0 & 0.1 & 17.09 &  88.13 & 66 & 1.34 & 0.04 & 2.1 \\
    0 & 0.01 & 0 & 0 & 1.0 & 16.12 &  110.38 & 66 & 1.67 & 0.0 & 3.48 \\
    0 & 0.01 & 0 & 0 & 10.0 & 16.01 &  124.3 & 66 & 1.88 & 0.0 & 4.28 \\
    0 & 0.01 & 0 & 0 & 100.0 & 16.0 &  136.06 & 66 & 2.06 & 0.0 & 4.92 \\
    0 & 0.1 & 0 & 0 & 0.001 & 27.88 &  48.08 & 66 & 0.73 & 0.95 & 0.06 \\
    0 & 0.1 & 0 & 0 & 0.01 & 26.91 &  52.17 & 66 & 0.79 & 0.89 & 0.13 \\
    0 & 0.1 & 0 & 0 & 0.1 & 22.0 &  64.57 & 66 & 0.98 & 0.53 & 0.63 \\
    0 & 0.1 & 0 & 0 & 1.0 & 17.09 &  100.46 & 66 & 1.52 & 0.0 & 2.88 \\
    0 & 0.1 & 0 & 0 & 10.0 & 16.12 &  122.74 & 66 & 1.86 & 0.0 & 4.19 \\
    0 & 0.1 & 0 & 0 & 100.0 & 16.01 &  135.7 & 66 & 2.06 & 0.0 & 4.9 \\
    0 & 1.0 & 0 & 0 & 0.001 & 27.99 &  45.67 & 66 & 0.69 & 0.97 & 0.03 \\
    0 & 1.0 & 0 & 0 & 0.01 & 27.88 &  49.34 & 66 & 0.75 & 0.94 & 0.08 \\
    0 & 1.0 & 0 & 0 & 0.1 & 26.91 &  54.08 & 66 & 0.82 & 0.85 & 0.19 \\
    0 & 1.0 & 0 & 0 & 1.0 & 22.0 &  70.18 & 66 & 1.06 & 0.34 & 0.96 \\
    0 & 1.0 & 0 & 0 & 10.0 & 17.09 &  110.12 & 66 & 1.67 & 0.0 & 3.46 \\
    0 & 1.0 & 0 & 0 & 100.0 & 16.12 &  132.86 & 66 & 2.01 & 0.0 & 4.74 \\
    0 & 10.0 & 0 & 0 & 0.001 & 28.0 &  44.0 & 66 & 0.67 & 0.98 & 0.02 \\
    0 & 10.0 & 0 & 0 & 0.01 & 27.99 &  47.33 & 66 & 0.72 & 0.96 & 0.05 \\
    0 & 10.0 & 0 & 0 & 0.1 & 27.88 &  50.99 & 66 & 0.77 & 0.91 & 0.11 \\
    0 & 10.0 & 0 & 0 & 1.0 & 26.91 &  56.23 & 66 & 0.85 & 0.8 & 0.25 \\
    0 & 10.0 & 0 & 0 & 10.0 & 22.0 &  73.77 & 66 & 1.12 & 0.24 & 1.18 \\
    0 & 10.0 & 0 & 0 & 100.0 & 17.09 &  116.81 & 66 & 1.77 & 0.0 & 3.85 \\
    0 & 100.0 & 0 & 0 & 0.001 & 28.0 &  46.78 & 66 & 0.71 & 0.96 & 0.04 \\
    0 & 100.0 & 0 & 0 & 0.01 & 28.0 &  46.44 & 66 & 0.7 & 0.97 & 0.04 \\
    0 & 100.0 & 0 & 0 & 0.1 & 27.99 &  49.34 & 66 & 0.75 & 0.94 & 0.08 \\
    0 & 100.0 & 0 & 0 & 1.0 & 27.88 &  52.98 & 66 & 0.8 & 0.88 & 0.16 \\
    0 & 100.0 & 0 & 0 & 10.0 & 26.91 &  58.09 & 66 & 0.88 & 0.75 & 0.33 \\
    0 & 100.0 & 0 & 0 & 100.0 & 22.0 &  76.68 & 66 & 1.16 & 0.17 & 1.36 \\
    \hline
    \enddata
\tablecomments{Espinoza reduction, decontaminated Visits 1-4 combined. $^\dagger$ Models with a N$\sigma$ equal to ``inf" do not agree with the data and are unambiguously ruled out.}
\end{deluxetable*}

\section{Forward Modeling and Retrieval Configurations}\label{sec:appendixD}

{\tt pRT} is used for forward modeling and {\tt POSIEDON} is used for retrievals. The opacity tables used for the codes are shown in Table \ref{tab:model_set_up}. Relevant collision-induced absorption (CIA) sources (e.g., N$_2$-N$_2$, O$_2$-O$_2$, N$_2$-O$_2$, N$_2$-H$_2$, H$_2$-H$_2$) and Rayleigh species (e.g., N$_2$, O$_2$, H$_2$) are used for each model \citep{Molliere2019prt, MacDonald2022}. The complete retrieval framework can be found in \url{https://github.com/nespinoza/TRAPPIST-1~e-GTO-2025}.

\begin{deluxetable}{lll}[h!]
    \renewcommand{\arraystretch}{1.1}
    \tabletypesize{\footnotesize}
    \tablecolumns{4} 
    \tablecaption{Forward Modeling and Retrieval Chemical Inventory}
    \tablehead{Molecule \phantom{space} & pRT Opacity References\phantom{space} & POSEIDON Opacity References\phantom{space}}
    \startdata
    \hline
    CO$_2$ & \citet{Yurchenko2020} & \citet{Yurchenko2020}\\
    CH$_4$ & \citet{Yurchenko2017} & \citet{Yurchenko2024}\\
    H$_2$O & \citet{Rothman2010} & \citet{Polyansky2018} \\
    CO & \citet{Rothman2010} & \citet{Li2015} \\
    N$_2$ & N/A & \citet{Gordon2017} \\
    O$_2$ & \citet{Gordon2017} & \citet{Gordon2017} \\
    O$_3$ & \citet{Rothman2013} & \citet{Gordon2017} \\
    N$_2$O & N/A & \citet{Gordon2017} \\
    C$_2$H$_2$ & \citet{Chubb2020} & N/A \\
    C$_2$H$_4$ & \citet{Mant2018} & N/A \\
    H$_2$S & \citet{Azzam2016} & N/A \\
    HCN & \citet{Harris2006, Barber2014} & N/A \\
    NH$_3$ & \citet{Yurchenko2011}  & N/A \\
    O & \citet{Kurucz1995} & N/A \\
    OH & \citet{Brooke2016} & N/A \\
    \enddata
    \label{tab:model_set_up}
    \tablecomments{Chemical inventory used in the forward modeling and retrieval analysis.}
\end{deluxetable}

\section{Additional Retrieval Constraints on high-$\mu$ Atmospheres}\label{sec:otheratms} \label{sec:appendixE}

\begin{figure*}[!ht]
    \centering
    \includegraphics[width=\textwidth]{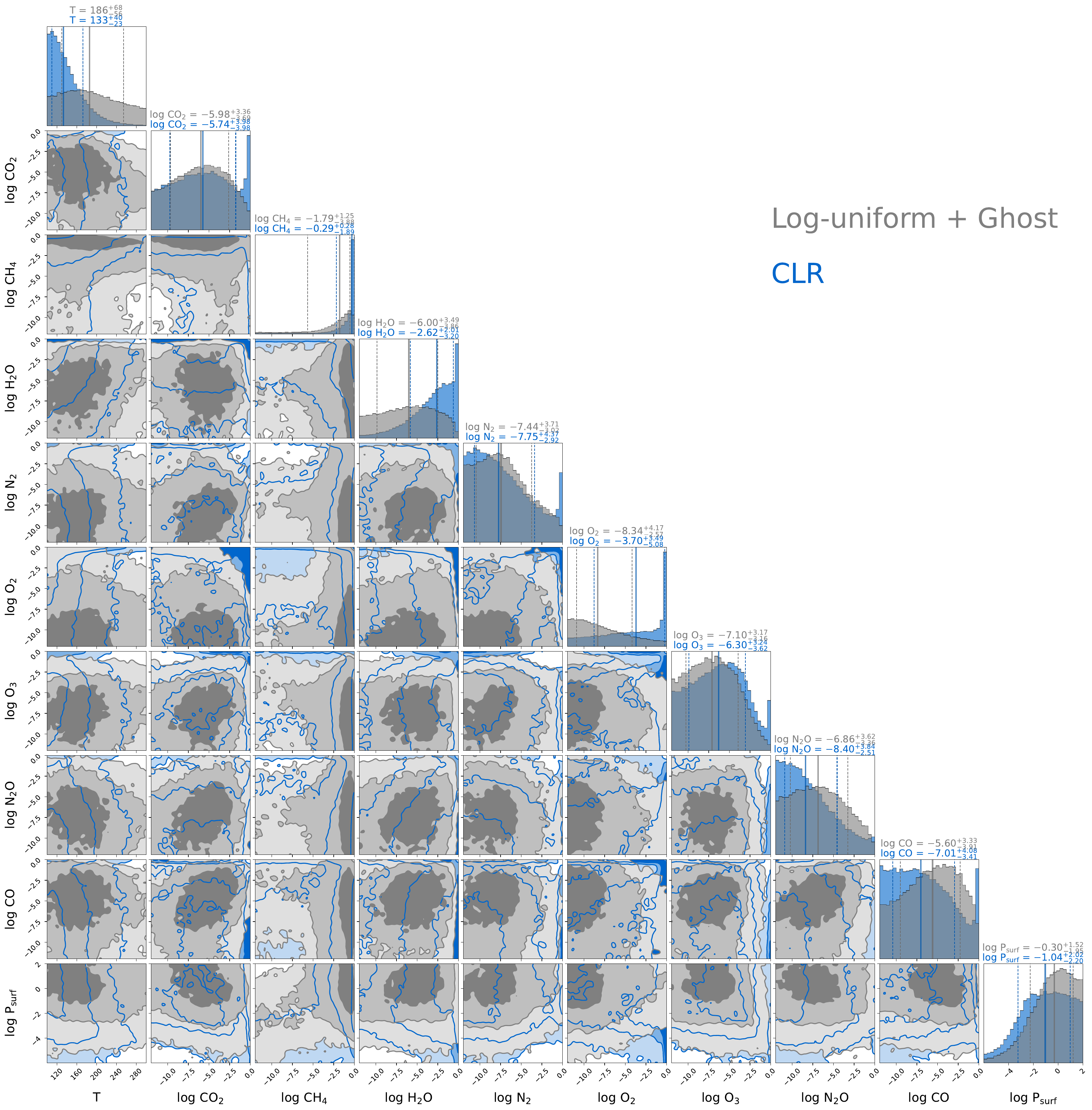}
    \caption{\textbf{Corner plot of both the CLR prior retrieval (blue) and log uniform prior with ghost background gas.} Parameters are listed on the y and x-axis for gas VMRs, effective surface pressure, and temperature. Posteriors are shown for each combination. Histograms are shown along the right edge. No molecules are detected.}
    \label{fig:corner}
\end{figure*}

In addition to CO$_2$ and CH$_4$, there are other spectrally active and inactive gases that we can constrain with our data. In Figure~\ref{fig:corner}, we show the posterior distributions for atmospheric properties (temperature, mixing ratios, and surface pressure) from our retrievals for CO, N$_2$O, O$_3$, O$_2$, N$_2$, H$_2$O, CH$_4$, and CO$_2$. In particular, we highlight that thick O$_3$-dominated atmospheres are ruled out ($P_{\rm{surf}} \lesssim 10^{-2}$\,bar to 2$\sigma$) because our observations cover the region near 4.8~$\micron$ where O$_3$ possesses a strong absorption feature. Additionally, high abundances for N$_2$O and CO are all disfavored to $>$2$\sigma$ (for the log-uniform priors + ghost gas retrieval) for both thin and thick atmospheres. The CLR prior allows pure compositions for all of the considered gases. For both the CLR and ghost gas retrieval, CH$_4$ is the only molecule consistently identified as a potential atmospheric component of TRAPPIST-1~e though it remains below the threshold of a detection.

\section{Retrieved Pressure-Temperature Posteriors Compared with the CO$_2$ Phase Diagram}\label{sec:co2phasediagram} \label{sec:appendixF}

\begin{figure*}[!ht]
    \centering
    \includegraphics[width=\textwidth]{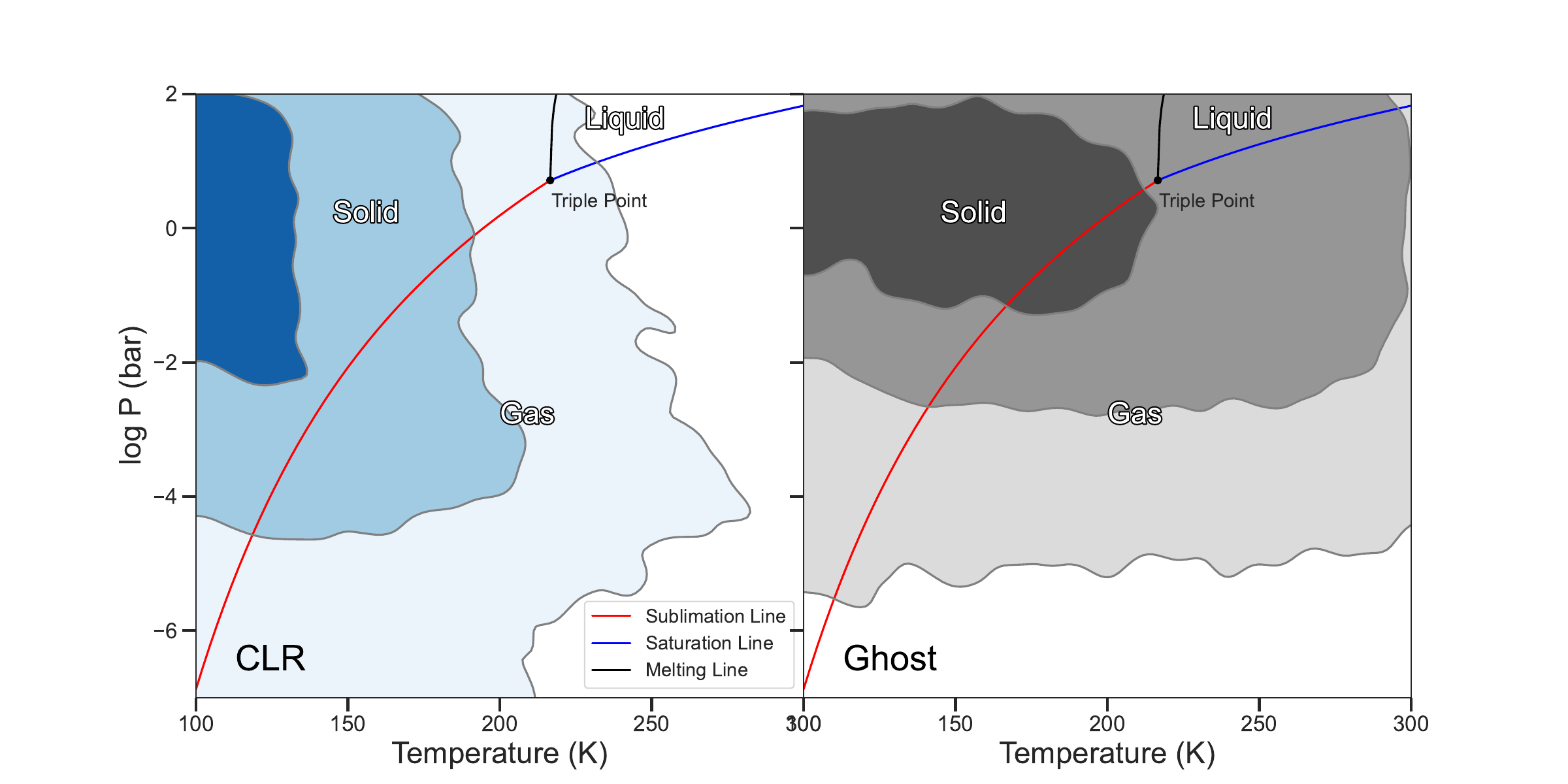}
    \caption{\textbf{Posterior for pressure-temperature compared with the CO$_2$ phase diagram.} Pressure in bars is on the y-axis and temperature in kelvins on the x-axis. The CLR posterior is shown on the left and the ghost on the right. Both the CLR and ghost retrieval find favorable regions of parameter space at cold temperatures where spectral features are minimized. However, favorable fits to the data include temperatures where CO$_2$ would condense out of the atmosphere, a process that is not accounted for by the retrieval.}
    \label{fig:phasediagram}
\end{figure*}

We compare our retrieved posteriors for pressure and temperature with the phase diagram for CO$_2$. The retrievals find good fits to the data by reducing the temperature and thereby flattening spectral features. In Figure \ref{fig:phasediagram} we show how the posterior space corresponds to the phase diagram for CO$_2$.

\end{document}